\def\ignore#1{}

\def\rn{}
\def\nn#1 #2{#2. #1}				
\def\nnn#1 #2 #3{#2. #3. #1}			
\def\nnnn#1 #2 #3 #4{#2. #3. #4 #1}		
\def\nnnnn#1 #2 #3 #4 #5{#2. #3. #4 #5. #1}	
\def\dualand{ and\hbox{ }}				
\def\multiand{, and\hbox{ }}				
\def\rf#1;#2;#3;#4;#5 {{\frenchspacing\par\rn#1, #3 {\bf #4}, #5 (#2). \par}}
\def\rg#1;#2;#3;#4;#5;#6 {{\frenchspacing\par\rn#1, #3 {\bf #4}, #5 (#2). \par}}
\def\rfbook#1;#2;#3;#4;#5 {{\frenchspacing\par\rn#1, {\it #3} (#5, #4, #2).\par}}
\def\rfprep#1;#2;#3 {{\par\frenchspacing\rn#1, #3 (#2).\par}}
\def\rfproc#1;#2;#3;#4;#5;#6 {{\frenchspacing\par\rn#1 #2, in {\it #3}, ed. #4 (#5: #6)\par}}
\def\rfprocp#1;#2;#3;#4;#5;#6;#7 {{\frenchspacing\par\rn#1 #2, in {\it #3}, ed. #4 (#5: #6), p#7\par}}

\def\beq#1{\begin{equation}\label{#1}}
\def\eeq{\end{equation}}
\def\beqa#1{\begin{eqnarray}\label{#1}}
\def\eeqa{\end{eqnarray}}
\def\eq#1{equation~(\ref{#1})}

\def\eqn#1{~(\ref{#1})}

\def\spose#1{\hbox to 0pt{#1\hss}}
\def\simlt{\mathrel{\spose{\lower 3pt\hbox{$\mathchar"218$}} \raise 2.0pt\hbox{$\mathchar"13C$}}}
\def\simgt{\mathrel{\spose{\lower 3pt\hbox{$\mathchar"218$}} \raise 2.0pt\hbox{$\mathchar"13E$}}}
\def\simpropto{\mathrel{\spose{\lower 3pt\hbox{$\mathchar"218$}} \raise 2.0pt\hbox{$\propto$}}}

\def\bt{\begin{tabbing}}
\def\et{\end{tabbing}}
\def\beq#1{\begin{equation}\label{#1}}
\def\eeq{\end{equation}}

\def\Sec#1{Section~\ref{#1}}

\def\bfig{\begin{figure}[h] \centerline{\hbox{}}\vfill}
\def\efig{\end{figure}\vfill\newpage}

\def\fig#1{Figure~\ref{#1}}

\def\fig#1{Figure~\ref{#1}}
\def\Fig#1{Figure~\ref{#1}}

\def\Cl{C_\l}

\def\lmax{{\ell_{\rm max}}}
\def\expec#1{\langle#1\rangle}
\def\ith{i^{th}}

\def\l{\ell}
\def\etal{{\frenchspacing\it et al.}}
\def\ie  {{\frenchspacing\it i.e.}}
\def\eg  {{\frenchspacing\it e.g.}}
\def\etc {{\frenchspacing\it etc.}}
\def\rms{rms}

\def\a{{\bf a}}
\def\c{{\bf c}}
\def\e{{\bf e}}

\def\n{{\bf n}}
\def\r{{\bf r}}
\def\rh{\hat{\r}}
\def\w{{\bf w}}
\def\x{{\bf x}}
\def\y{{\bf y}}

\def\xt{\tilde{\x}}

\def\rh{\widehat{\r}}
\def\A{{\bf A}}

\def\C{{\bf C}}
\def\Clcmb{C_\l^{\rm cmb}}
\def\CClcmb{\C_\l^{\rm cmb}}
\def\Cljunk{\C_\l^{\rm junk}}
\def\Clclean{C_\l^{\rm clean}}
\def\Clnocmb{\C_\l^{\rm nocmb}}
\def\Ct{{\bf\tilde C}}

\def\F{{\bf F}}

\def\I{{\bf I}}
\def\M{{\bf M}}
\def\N{{\bf N}}

\def\L{{\bf L}}

\def\T{{\bf T}}

\def\W{{\bf W}}
\def\PP{{\bf\Pi}}

\def\q{{\bf q}}

\def\Cl{C_\l}

\def\ith{i^{th}}

\def\bs{}

\def\mK{{\rm mK}}
\def\uK{\mu{\rm K}}

\def\alm{a_{\ell m}}
\def\Ylm{Y_{\ell m}}

\def\lbar{{\bar\l}}
\def\Ct{{\tilde C}}
\def\Ctot{C^{\rm tot}}
\def\Ccmb{C^{\rm cmb}}
\def\Cnoise{C^{\rm noise}}

\def\figsize{9.0cm}

\documentclass[twocolumn,amsmath,nofootinbib]{revtex4}
\usepackage{url}
\begin{document}

\input{epsf.sty}

\title{A high resolution foreground cleaned CMB map from WMAP}

\author{Max Tegmark$^{1}$, Ang\'elica de Oliveira-Costa$^{1}$ \& Andrew J. S. Hamilton$^2$}      
\address{$^{1}$Department of Physics \& Astronomy, University of Pennsylvania, Philadelphia, PA 19104, USA, max@physics.upenn.edu\\}
\address{$^{2}$JILA and Dept\. of Astrophysical and Planetary Sciences, U. Colorado, Boulder, CO 80309, USA, Andrew.Hamilton@colorado.edu}

\date{Submitted to Phys. Rev. D March 4 2003, accepted July 25.}


\begin{abstract}
We perform an independent foreground analysis of the WMAP maps to produce a cleaned
CMB map (available online) useful for cross-correlation with, \eg,  galaxy and
X-ray maps. We use a variant of the Tegmark \& Efstathiou (1996) technique that
assumes that the CMB has a blackbody spectrum,
but is otherwise
completely blind, making no assumptions about the CMB
power spectrum,
the foregrounds, WMAP detector noise or external templates. Compared with the
foreground-cleaned internal linear combination map produced by the WMAP team, our
map has the advantage of containing less non-CMB power (from foregrounds and
detector noise) outside the Galactic plane. The difference is most important on 
the the angular scale of the first acoustic peak and
below,
since our cleaned map is at the highest (12.6$^\prime$) rather
than lowest (49.2$^\prime$) WMAP resolution. We also produce a Wiener filtered CMB map,
representing our best guess as to what the CMB sky actually looks like, as well as
CMB-free maps at the five WMAP frequencies useful for foreground studies.

We argue that our CMB map is clean enough that the lowest multipoles can be measured
without any galaxy cut, and obtain a quadrupole value that is slightly less low than that
from the cut-sky WMAP team analysis. This can be understood from a map of the CMB
quadrupole, which shows much of its power falling within the Galaxy cut region,
seemingly coincidentally. Intriguingly, both the quadrupole and the octopole are seen
to have power suppressed along a particular spatial axis, which lines up between the two,
roughly towards $(l,b)\sim (-110^\circ,60^\circ)$ in Virgo.

\bigskip
\end{abstract}

  

\maketitle

\section{INTRODUCTION}

The release of the first results 
\cite{BennettWMAP,BennettMission,BennettForegs,barnes1-03,barnes2-03,hinshawpower03,hinshaw03,jarosik1-03,jarosik2-03,kogut03,komatsu03,limon03,page1-03,page2-03,Page3-03,peiris03,Spergel03,verde03}
from the Wilkinson Microwave Anisotropy Probe (WMAP)
constituted a major milestone in cosmology, laying a solid foundation upon which to found
the cosmological quest in coming years.
Although much of the attention in the wake of the WMAP release has focused on ``sexy'' 
issues like the power spectrum and its cosmological implications, 
the primary stated science goal of WMAP is to produce maps.
Indeed, one of the qualitatively new types of research made possible by WMAP involves taking advantage
of this spatial information by cross-correlating the maps with other cosmological templates such as
galaxy \cite{Peiris00}, x-ray \cite{Boughn02,DiegoSilk03}, infrared and lensing maps \cite{Song02}, 
which can reveal interesting signals ranging from the Late Integrated Sachs-Wolfe effect to 
the SZ effect and lensing \cite{Cooray01,Hirata02}.

Many such future studies will be looking for signals of modest statistical significance, 
so it is important to quantify and minimize unwanted signals in the map due to foreground contamination
and detector noise. Accurately understanding the foreground signal is also important for the 
interpretation of the WMAP early reionization detection
\cite{BennettWMAP,Page3-03,Spergel03,Haiman03,Holder03,Hui03}
and for the interpretation of the low WMAP quadrupole
\cite{BennettWMAP,hinshawpower03,Spergel03},
since Galactic foregrounds are most important on large angular scales
\cite{wiener,polarforgs}.
The WMAP team has already performed a careful foreground analysis \cite{BennettForegs} 
combining the five frequency bands into a single foreground-cleaned map, shown in 
Figure 1 (top).
Given the huge effort that has gone into creating the spectacular multifrequency maps, 
it is clearly worthwhile to subject them to an independent foreground analysis.
This is the purpose of the present paper, which we will argue not only corroborates the findings of the WMAP team,
but also makes some further improvements that we believe are useful.

The main goal of this paper is to remove foregrounds, not to understand or model them.
For reviews of foreground modeling issues, see, \eg, 
\cite{BennettForegs,wiener,foregpars,polarforgs,forgbook,forgx,Bouchet1,Bouchet2,Bouchet3,tucci00,Baccigalupi00,Burigana02,bruscoli02,tucci02,giardino02} and references therein.
There is a rich literature of techniques for foreground removal. 
The work most closely related to the present paper is that
done in preparation for the Planck mission \cite{Bouchet1,Bouchet2,Bouchet3},
developing multipole-based cleaning techniques and testing them on simulations.

\begin{figure*} 

\centerline{\epsfxsize=15.1cm\epsffile{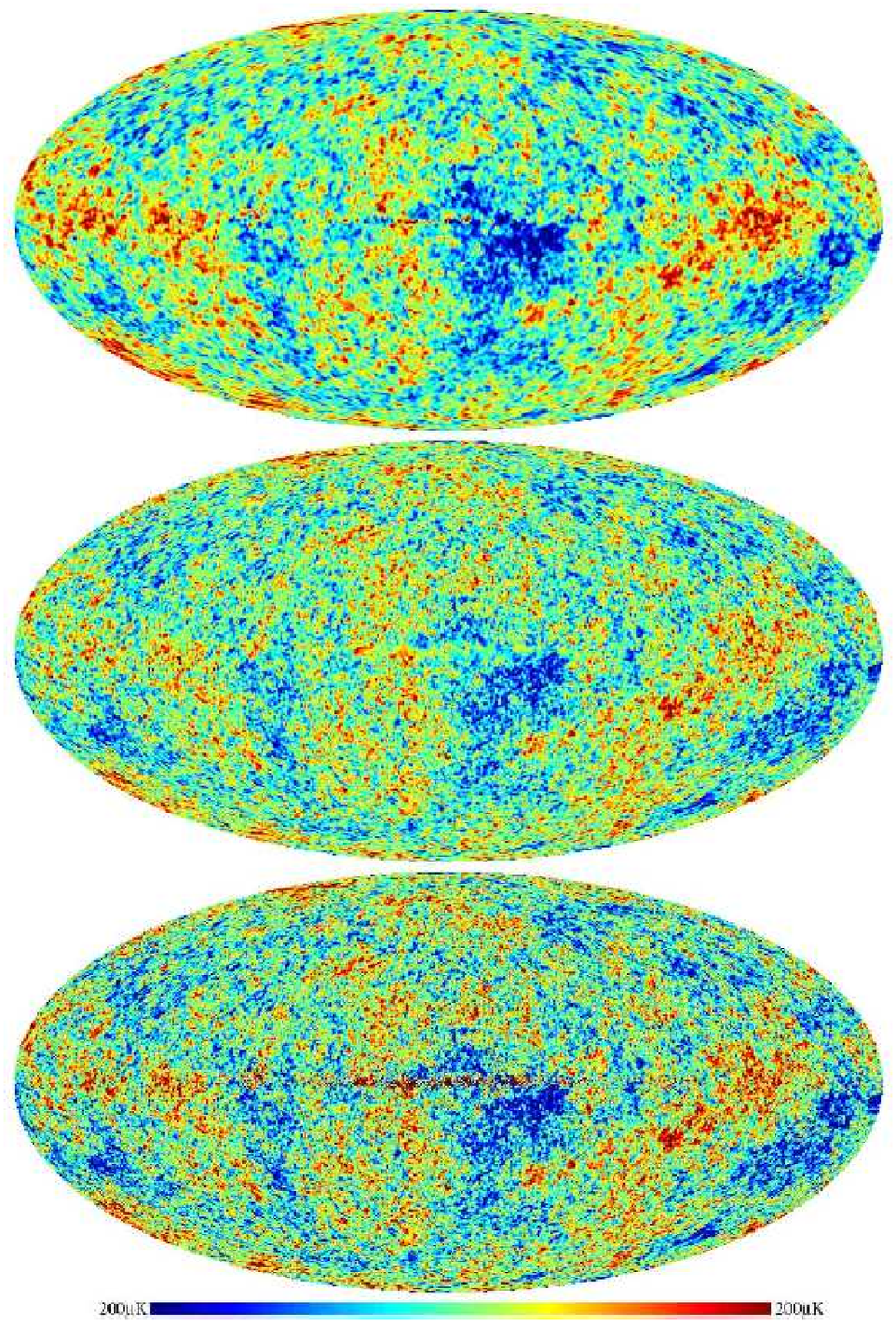}}
\vskip-26.05cm
\centerline{\epsfxsize=14.7cm\epsffile{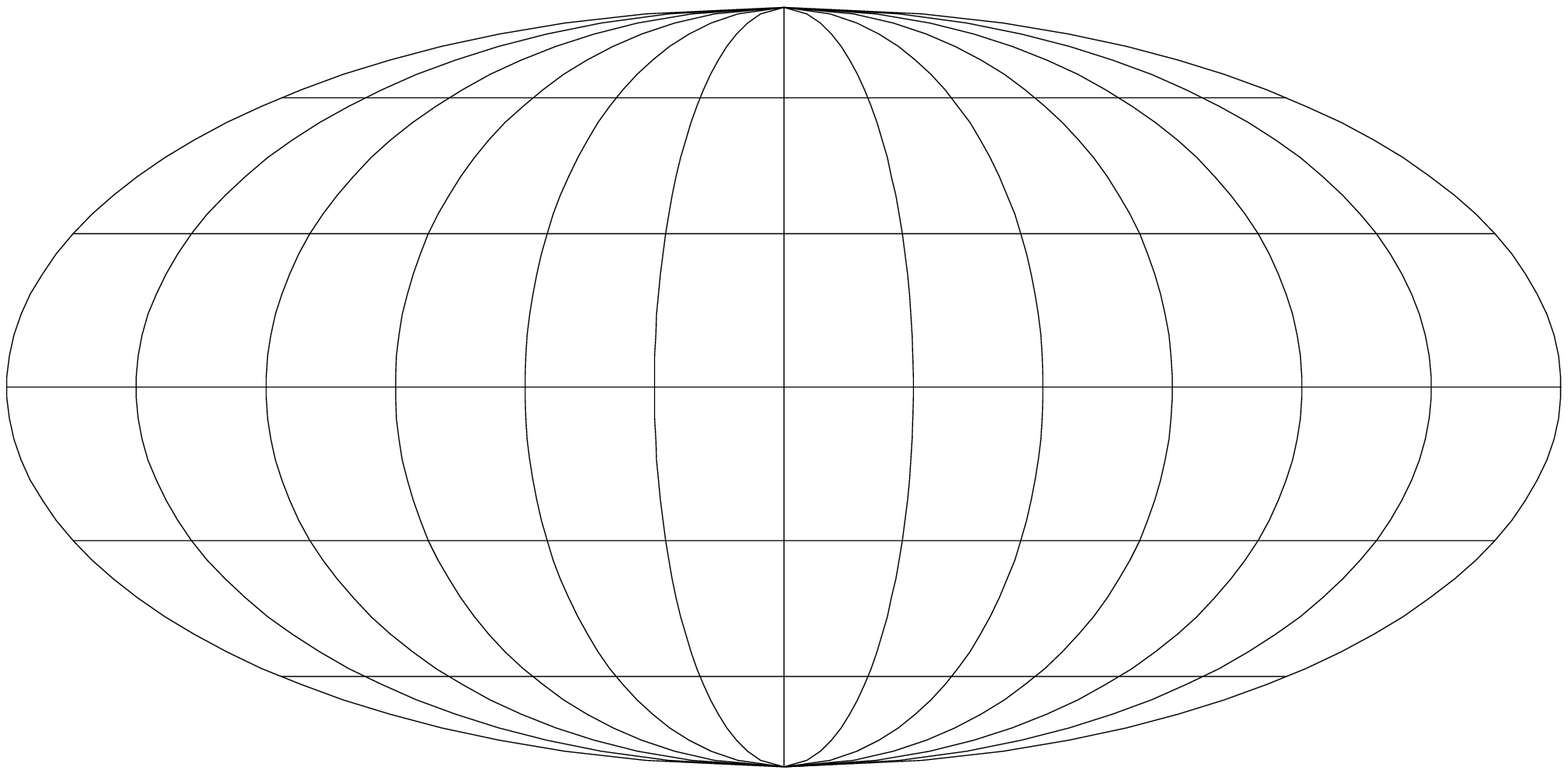}}
\vskip-7.42cm
\centerline{\epsfxsize=14.7cm\epsffile{mollweide_lores.ps}}
\vskip-7.42cm
\centerline{\epsfxsize=14.7cm\epsffile{mollweide_lores.ps}}
\vskip-3cm


\noindent
\parbox[l]{18.0cm}{\footnotesize\flushleft
FIG.~1.~The linearly cleaned WMAP team map (top), 
our Wiener filtered map (middle) and our raw map (bottom).
Here and throughout, all maps are shown in Mollweide projection
in Galactic coordinates with the Galactic center $(l,b)=(0,0)$ in the middle and 
Galactic longitude $l$ increasing to the left.
These low-resolution
images do not do justice to the WMAP data or our method; 
full res versions are available at $http://www.hep.upenn.edu/~max/wmap.html$.
}
\end{figure*}

\begin{figure*} 
\hglue0cm\epsfxsize=17.5cm\epsffile{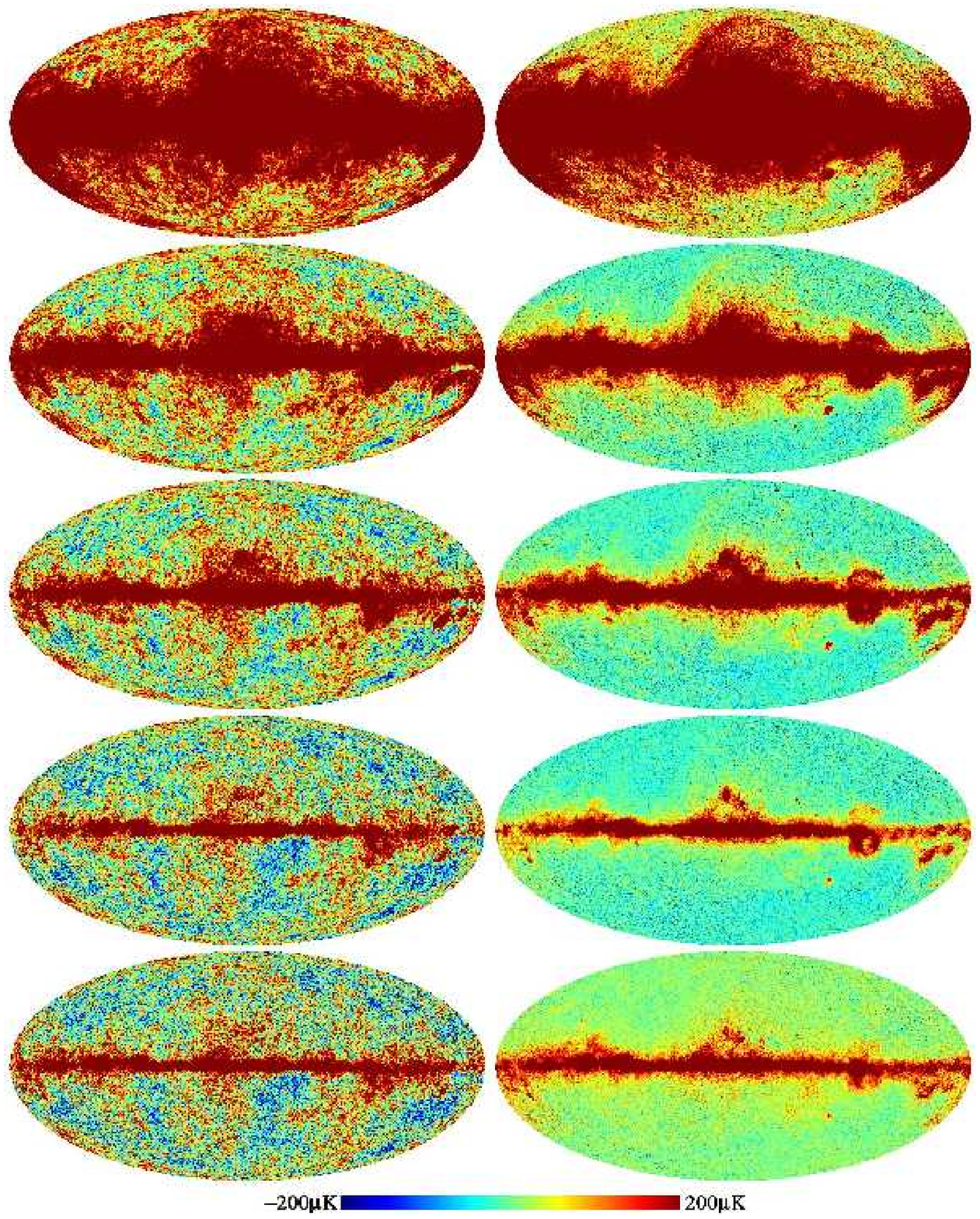}

\smallskip

\noindent
\hbox{\footnotesize FIG.~2.~The five WMAP frequency bands
K, Ka, Q, V and W
(top to bottom)
before (left) and after (right) removing the CMB.
}
\end{figure*}

\section{ANALYSIS AND RESULTS}

\subsection{The problem}

A key goal of the CMB community is to measure the function
$x(\rh)$, the true CMB sky temperature in the sky direction given by 
the unit vector $\rh$. The WMAP team have observed the sky 
in five frequency bands denoted K, Ka, Q, V and W, centered on the frequencies of
22.8, 33.0, 40.7, 60.8 and 93.5 GHz, respectively, producing five corresponding maps 
(Figure 2, left)
that we will refer to 
as $\y_i$, $i=1,...,5$. 
In practice, each of these maps are discretized into $n=12\times 512^2 = 3$,$145$,$728$ 
HEALPix\footnote{The HEALPix 
package is available from
\protect\url{http://www.eso.org/science/healpix/}.
}
pixels \cite{healpix1,healpix2}, 
so $\y_i$ is an $n$-dimensional vector. However, since the maps are 
more than adequately oversampled relative to the beam resolution, it is equivalent and often simpler to think
of them simply as five smooth
functions
$y_i(\rh)$. Conversely, we will occasionally write the true sky as a 
pixelized vector $\x$.

These maps are related to the true sky $\x$ by the affine relation
\beq{ModelEq1}
\y_i = \A_i\x+\n_i,
\eeq
where the matrix $\A_i$ encodes the effect of beam smoothing and the additive term $\n_i$
is the contribution from detector noise and foreground contamination, collectively referred to as ``junk'' below
since it complicates the recovery of $\x$.
Defining the larger matrices and vectors
\beq{GroupingEq}
\A\equiv\left(\bs\begin{tabular}{c}
$\A_1$\\
$\vdots$\\
$\A_5$
\end{tabular}\bs\right),\quad
\y\equiv\left(\bs\begin{tabular}{c}
$\y_1$\\
$\vdots$\\
$\y_5$
\end{tabular}\bs\right),\quad
\n\equiv\left(\bs\begin{tabular}{c}
$\n_1$\\
$\vdots$\\
$\n_5$
\end{tabular}\bs\right),\quad
\eeq
we can rewrite the system of equations given by\eqn{ModelEq1} as 
\beq{ModelEq2}
\y=\A\x+\n,
\eeq
a set of linear equations that would be highly over-determined by a factor five
if it were not for the presence of unknown junk
$\n$.

Foreground removal involves 
inverting this overdetermined system of noisy linear equations. 
The most general linear\footnote{In addition to simplicity and transparency, 
linear methods have the advantage that the noise properties 
of
$\xt$ can be readily calculated from those of the input maps.}
estimate of $\xt$ of the true sky $\x$ can be written 
\beq{WdefEq}
\xt = \W\y
\eeq
for some $n\times (5n)$ matrix $\W$.
We will require that the inversion leaves the true sky unaffected, \ie, 
that the expected measurement error $\expec{\xt}-\x$ is independent of $\x$.
Bennett {\etal} \cite{BennettForegs} refer to this property as the inversion having unit response to
the CMB. Methods involving maximum-entropy reconstruction or some form of smoothing typically lack this
property.
Substituting \eq{ModelEq2} into \eq{WdefEq} shows that this requirement corresponds to the constraint
\beq{WAeq}
\W\A=\I.
\eeq

\subsection{The mathematically optimal solution}

Which choice of $\W$ gives the smallest {\rms} errors from foregrounds and detector noise combined?
Physically different but mathematically identical problems 
were solved in a CMB context by \cite{wright96,tegmark97},
showing that the minimum-variance choice is
\beq{ComboEq1}
\xt = [\A^t\N^{-1}\A]^{-1}\A^t\N^{-1}\y,
\eeq
where $\N\equiv\expec{\n\n^t}$. For an extensive discussion of different methods
proposed in the literature and the relations between
them,
see \cite{foregpars}.
Although optimal, this method is unfortunately unfeasible for the WMAP case, for two reasons.
First, it requires the inversion of
the $(5n)\times (5n)$ matrix $\N$.
Although the detector noise contribution to this matrix is close to diagonal for WMAP, the foreground
contribution is certainly not.
Second, it requires knowing the matrix $\N$, which has many more components than there are pixels in the map.

\subsection{The WMAP team solution}

In producing their internal linear combination (ILC) map, 
the WMAP team adopt a simpler approach \cite{BennettForegs}, first smoothing all five maps to a common 
angular resolution of $1^\circ$ and then performing the cleaning separately for each pixel.
The smoothing implies that $\A_i=\I$ (if we redefine $\x$ to be the true sky
smoothed
to $1^\circ$), 
and \eq{WdefEq} reduces to the simple form 
\beq{BennettMethodEq}
\xt(\rh) = \sum w_i\y_i(\rh)
\eeq
for five weights $w_i$ that according to \eq{WAeq} must sum to unity.
The WMAP team chose the weights that minimize the rms fluctuations in the cleaned map $\xt$, 
using 12 separate weight vectors for 12 disjoint sky regions.

Although this method works well,
it can be improved by allowing the weights
to depend on angular scale (i.e.\ on harmonic number $\l$)
as well as on Galactic latitude.
This has two advantages.
First,
the angular resolution is limited not by
that of the lowest resolution channel
(the K-band FWHM is $49.2^\prime$),
but rather by 
that of the highest resolution channel
(the W-band FWHM is $12.6^\prime$).
Second, as shown by \cite{wiener},
letting the weights depend on angular scale
produces a cleaner map
by taking into account that the frequency
dependence of the unwanted signals varies with scale:
at large scales, galactic foregrounds dominate,
whereas at small scales, detector noise dominates.

\setcounter{figure}{2}

\begin{figure} 
\centerline{\epsfxsize=\figsize\epsffile{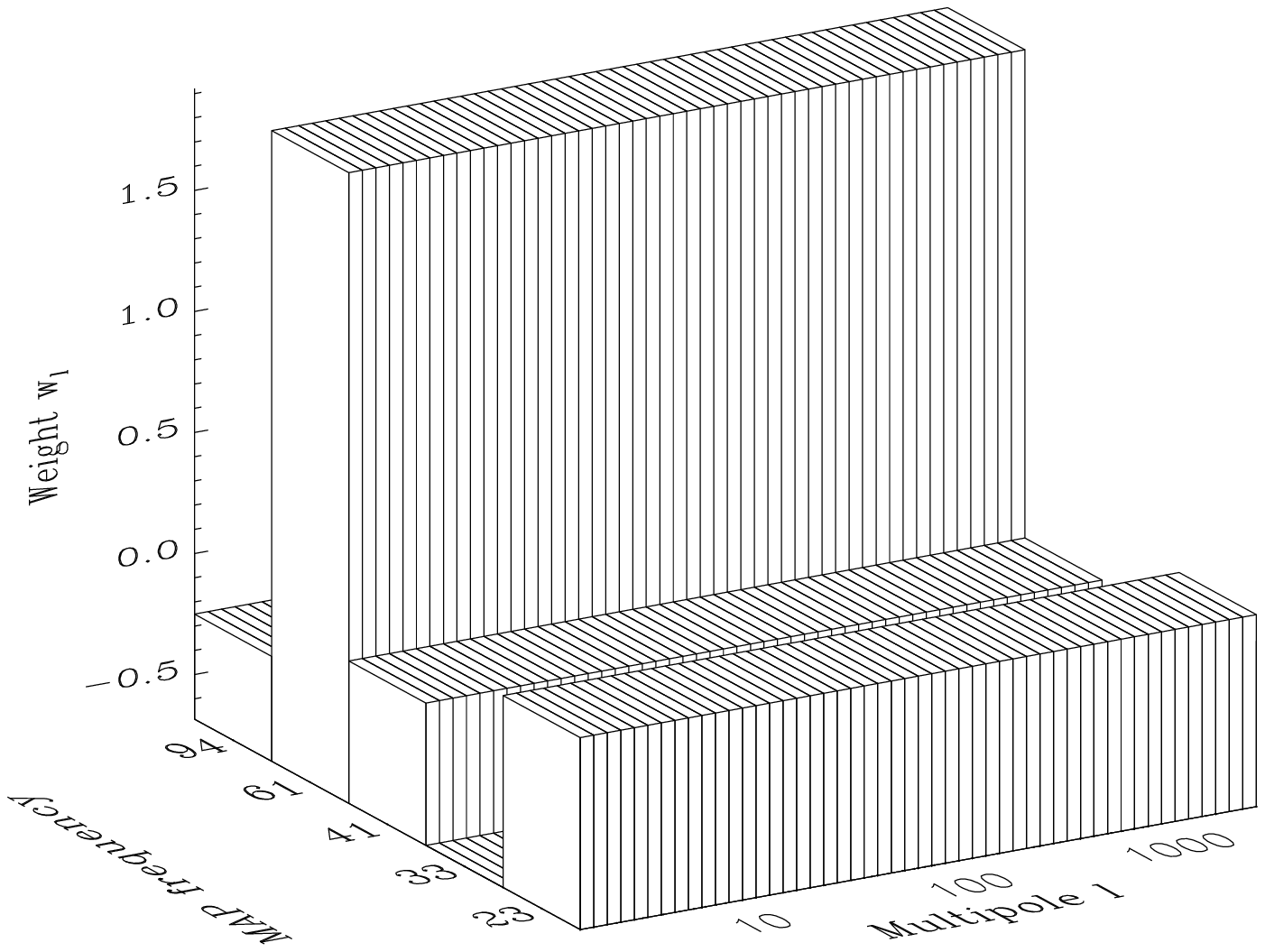}}
\caption[1]{\label{bennett_weightsFig}\footnotesize%
The weights $\w_\l$
used to create the internal linear combination map of the WMAP team
are independent of angular scale. The figure shows the
weights $\w_\l = (0.109, -0.684, -0.096, 1.921, -0.250)$ used outside of the galactic plane.
}
\end{figure}

\begin{figure} 
\centerline{\epsfxsize=\figsize\epsffile{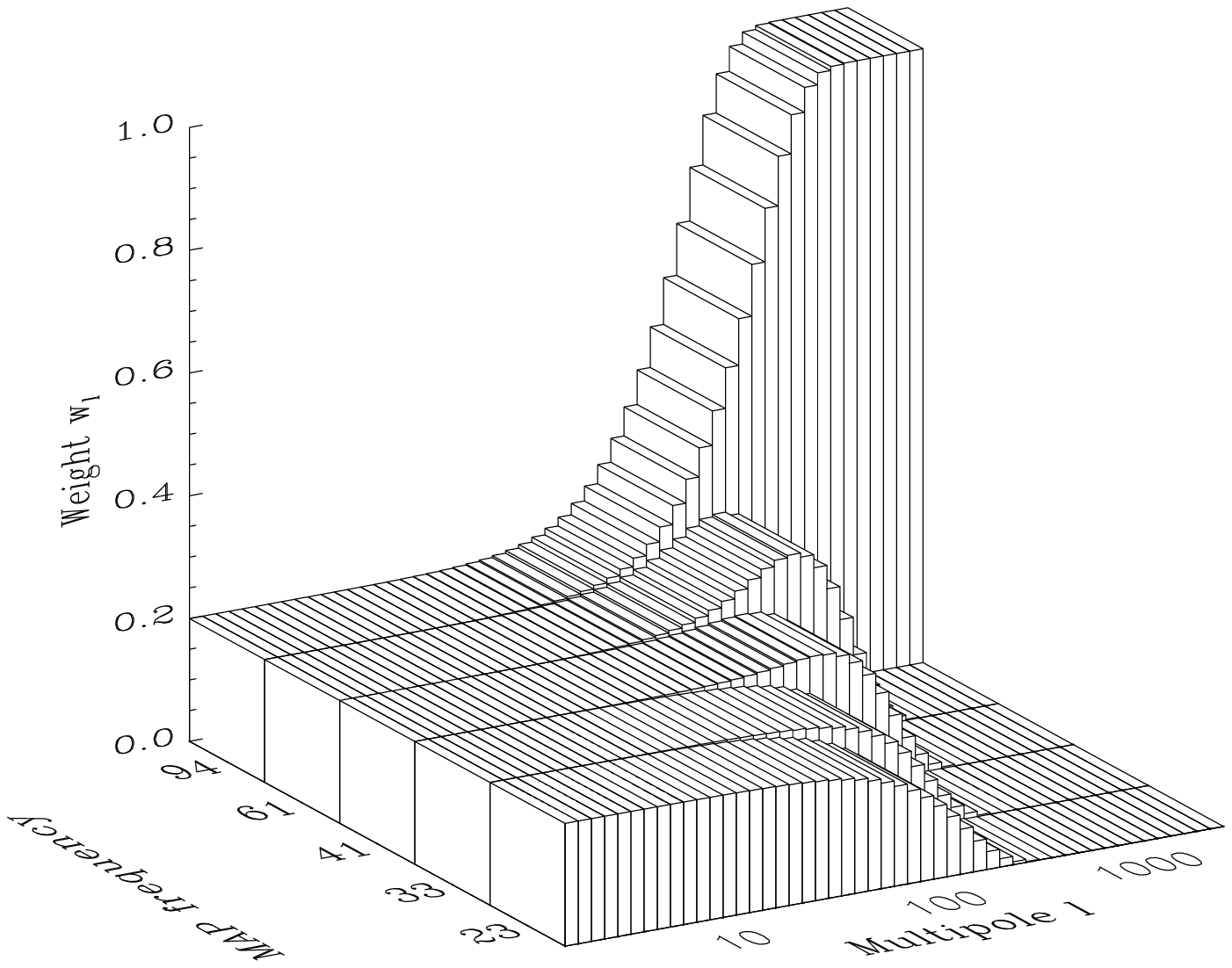}}
\caption[1]{\label{noforeg_weightsFig}\footnotesize%
If there were no foregrounds and equal noise in the five input maps,
then equal weighting at low $\l$ would give way to favoring the highest
resolution bands at high $\l$. This example uses the forecast WMAP beam and noise 
specifications from \protect\cite{foregpars} rather than the actual ones.
}
\end{figure}

\begin{figure} 
\centerline{\epsfxsize=\figsize\epsffile{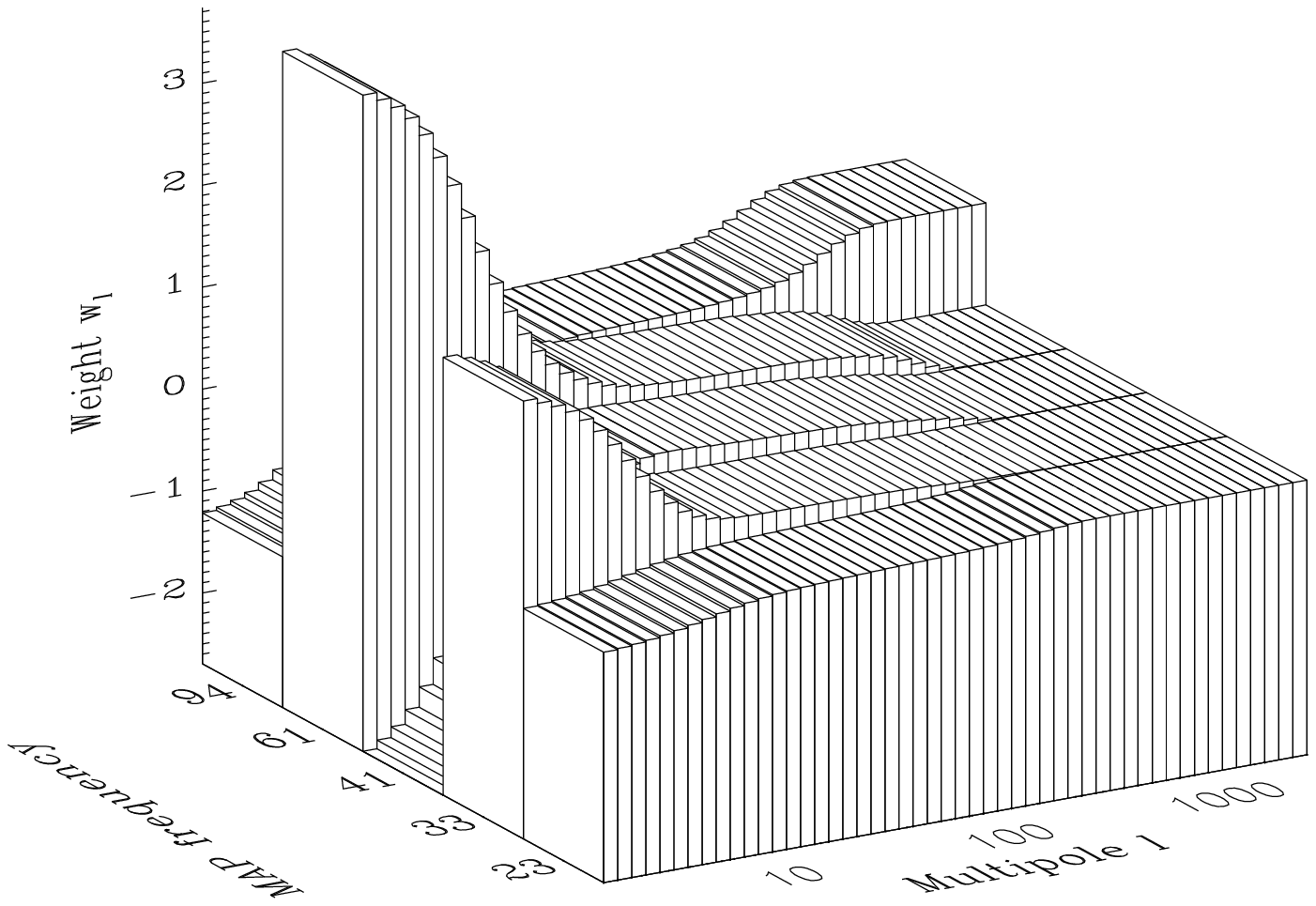}}
\caption[1]{\label{midforeg_weightsFig}\footnotesize%
The optimal WMAP weights forecast by \cite{foregpars} for the 
middle-of-the-road foreground model from \cite{foregpars}.
}
\end{figure}

\begin{figure} 
\centerline{\epsfxsize=\figsize\epsffile{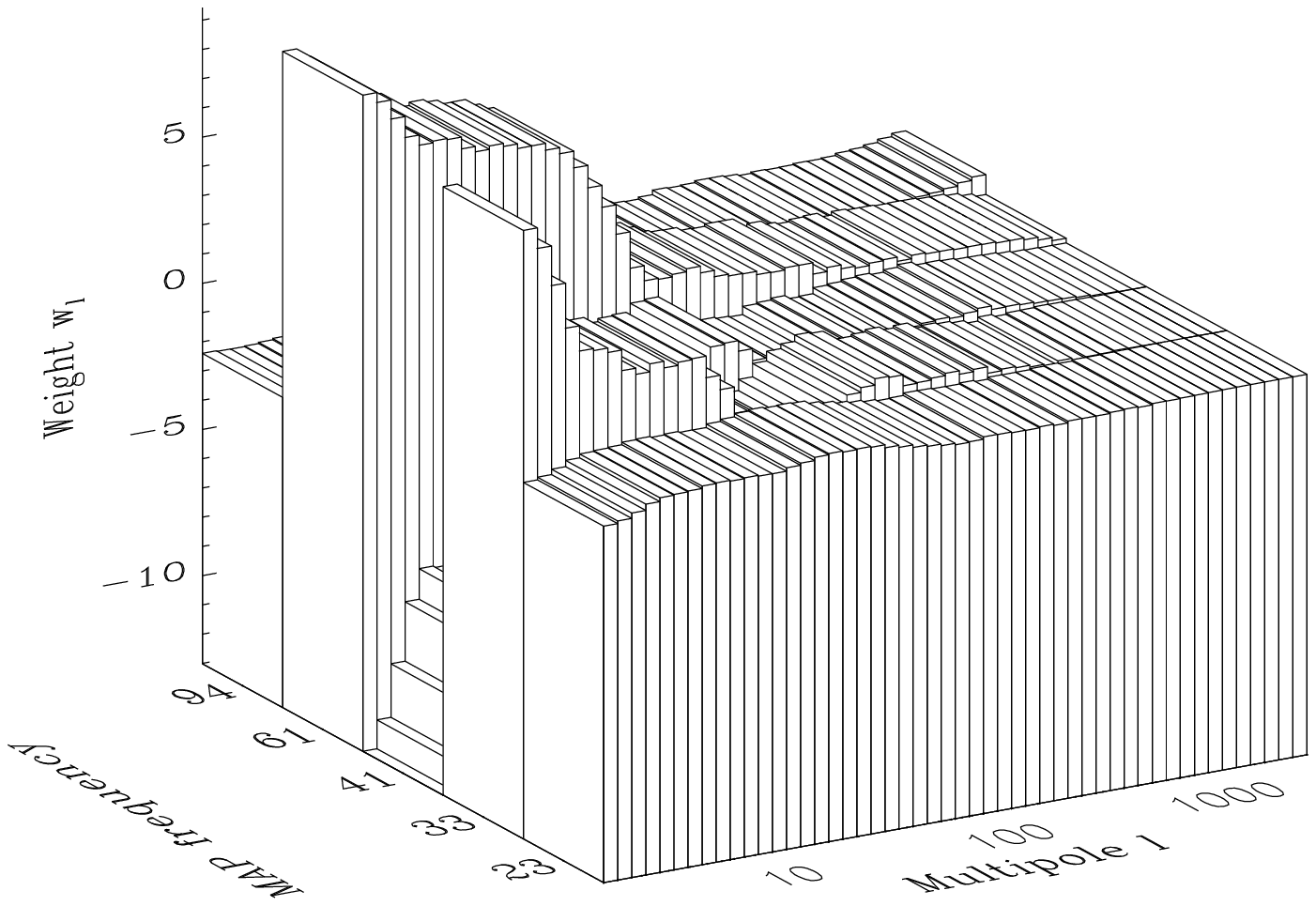}}
\caption[1]{\label{mask5foreg_weightsFig}\footnotesize%
The actual weights we use for the 3rd cleanest of the
9 sky regions
shown in \fig{masksFig}.
The dirtier the sky region, the more aggressive the weighting 
becomes, using large negative and positive values to subtract foregrounds.  
}
\end{figure}

\subsection{Our solution}

In this paper, we will take an approach intermediate between the two described above, 
aiming to approximate the optimal method while staying within the realm of numerical feasibility.
Our method is essentially that of Tegmark \& Efstathiou \cite{wiener}, implemented to make it blind and 
free of assumptions both about the CMB power spectrum and 
about foreground and noise properties. The only assumption, which is crucial,
is that the CMB has the Blackbody spectrum determined by the COBE/FIRAS experiment
\cite{Mather94}.
The method
strictly respects the requirement of \eq{WAeq}, which is most easily seen by verifying that each of the
steps described below do so individually. The gist of our method is to combine
the five WMAP bands with weights that depend both on angular scale and on distance to the Galactic plane
(we subdivide the sky into 9 regions of decreasing overall cleanliness).
We begin by describing the angular scale separation, then turn to the spatial subdivision in \Sec{MaskSec}.

We perform our cleaning in multipole space as in \cite{wiener}, 
and go back and forth between spherical harmonic coefficients 
\beq{almDefEq}
\alm^i\equiv\int \Ylm(\rh)^* \x_i(\rh)d\Omega
\eeq
and real-space maps
\beq{almInvEq}
x_i(\rh) = \sum\alm^i\Ylm(\rh)
\eeq
using the HEALPix package \cite{healpix1,healpix2} 
with $\lmax=1024$. 
Since this employs a spherical harmonic transform algorithm
using fast Fourier transforms in the azimuthal direction 
\cite{Muciaccia97}, 
each transformation 
takes only about a minute on a Linux workstation.
Each cleaned map is defined by 
\beq{CleanedalmEq}
\alm\equiv\sum_{i=1}^5 w_\l^i {\alm^i\over B_\l^i}
\eeq
for some set of five-dimensional weight vectors $\w_\l$, where
$B_\l^i$ is the beam function for the $\ith$ channel from \cite{page2-03}.
(There are 4 W-band maps, 2 V-band maps and 2 Q-band maps; we combine these
into single maps at each frequency by straight averaging and therefore average the
corresponding beam functions as well.)
When computing our final cleaned map in real space, we multiply $\alm^i$ by $B_\l^5$ in 
\eq{almInvEq} so that it has the beam corresponding to the highest-resolution WMAP band.

To gain intuition for the weight vectors $\w_\l$ that specify a cleaned map, 
we have plotted them for four interesting cases in figures~\ref{bennett_weightsFig},
\ref{noforeg_weightsFig},
\ref{midforeg_weightsFig} and
\ref{mask5foreg_weightsFig}.
\Fig{bennett_weightsFig} corresponds to the weighting used by the WMAP team for the
region away from the galactic plane,
and is simply independent of $\l$.
To recover their published internal
linear combination map shown in 
Figure 1 (top),
one simply
applies these weights after first multiplying each
$\alm^i$ by a Gaussian beam 
with FWHM=1$^\circ$ in \eq{almInvEq}.

The main drawback of this weighting is that it neglects that there is a 
tradeoff between foregrounds and detector noise which depends strongly on angular scale.
Diffuse foregrounds are most important on large scales where
detector
noise is negligible, warranting 
large negative and positive weights to aggressively subtract foregrounds.
Detector noise, on the other hand, is most important on small scales, both because 
of its Poissonian nature ($\C_l$ roughly constant in the observed map) and because 
the beam correction in \eq{CleanedalmEq} causes it to blow up exponentially on scales smaller than
the angular resolution \cite{KnoxNoise,wiener}.
In the limit $\l\to\infty$, the best weighting is therefore $\w_\l\to (0,0,0,0,1)$
regardless of what the foregrounds are doing, as illustrated in \fig{noforeg_weightsFig},
since the W-band has the best resolution.

We choose to minimize the total unwanted power from foregrounds and noise combined,
separately for each
harmonic
$\l$ as in \cite{wiener},
as opposed to only for the combination 
corresponding to the $1^\circ$
pixel variance
as in \cite{BennettForegs}.
As seen in \fig{midforeg_weightsFig}, one expects such a weighting to combine features from the two previous 
figures: rather aggressive subtraction using all five channels at low $\l$, more
cautious subtraction using only the higher-resolution channels on intermediate scales, and
all the weight on the W-band at extremely high $\l$. 
In particular,
it is crucial to downweight the K-band when cleaning on the scales of the acoustic peaks,
otherwise
the acoustic peaks in the resulting map will be dominated by K-band noise.

The constraint \eq{WAeq} that we leave the CMB untouched corresponds to the requirement that the weights sum
to unity
($\sum_i\w_\l^i=1$) for each $\l$, \ie, that
\beq{wsumEq}
\e\cdot\w_\l=1,
\eeq
where $\e=(1,1,1,1,1)$ is a column vector of all ones.
Minimizing the power $\expec{|\alm|^2}$ in the cleaned map of \eq{CleanedalmEq} subject to this constraint 
gives \cite{wiener,foregrounds,foregpars}
\beq{OptimalwEq}
\w_\l = {\C_\l^{-1}\e\over \e^t\C_\l^{-1}\e},
\eeq
where $\C_\l$ is the $5\times 5$ matrix-valued cross-power spectrum
\beq{CmatrixEq}
\C_\l^{ij}\equiv\expec{{\alm^i}^*\alm^j}.
\eeq
As an example, \fig{mask5foreg_weightsFig} shows the weights we obtain for the 3rd cleanest of the
9
sky regions shown in \fig{masksFig} below. We see that just as
forecast in \fig{midforeg_weightsFig},
and as in the WMAP team weighting of \fig{bennett_weightsFig},
the 61 GHz
V
channel is ``the breadwinner'', getting a large positive weight on large scales since it has the
lowest overall foreground level. The 94 GHz
W channel gets a negative weight to subtract out dust,
and the three lower frequency channels are used to subtract out synchrotron, free-free and any other emission
dominating at low frequencies.
In cleaner sky regions,
weights get less aggressive in the sense of acquiring 
smaller absolute values. In particular, we recover weights similar to those of the WMAP team
(\fig{bennett_weightsFig}) on large scales for the Kp2 sky cut defined and used by
\cite{BennettForegs}.

\begin{figure} 
\centerline{\epsfxsize=\figsize\epsffile{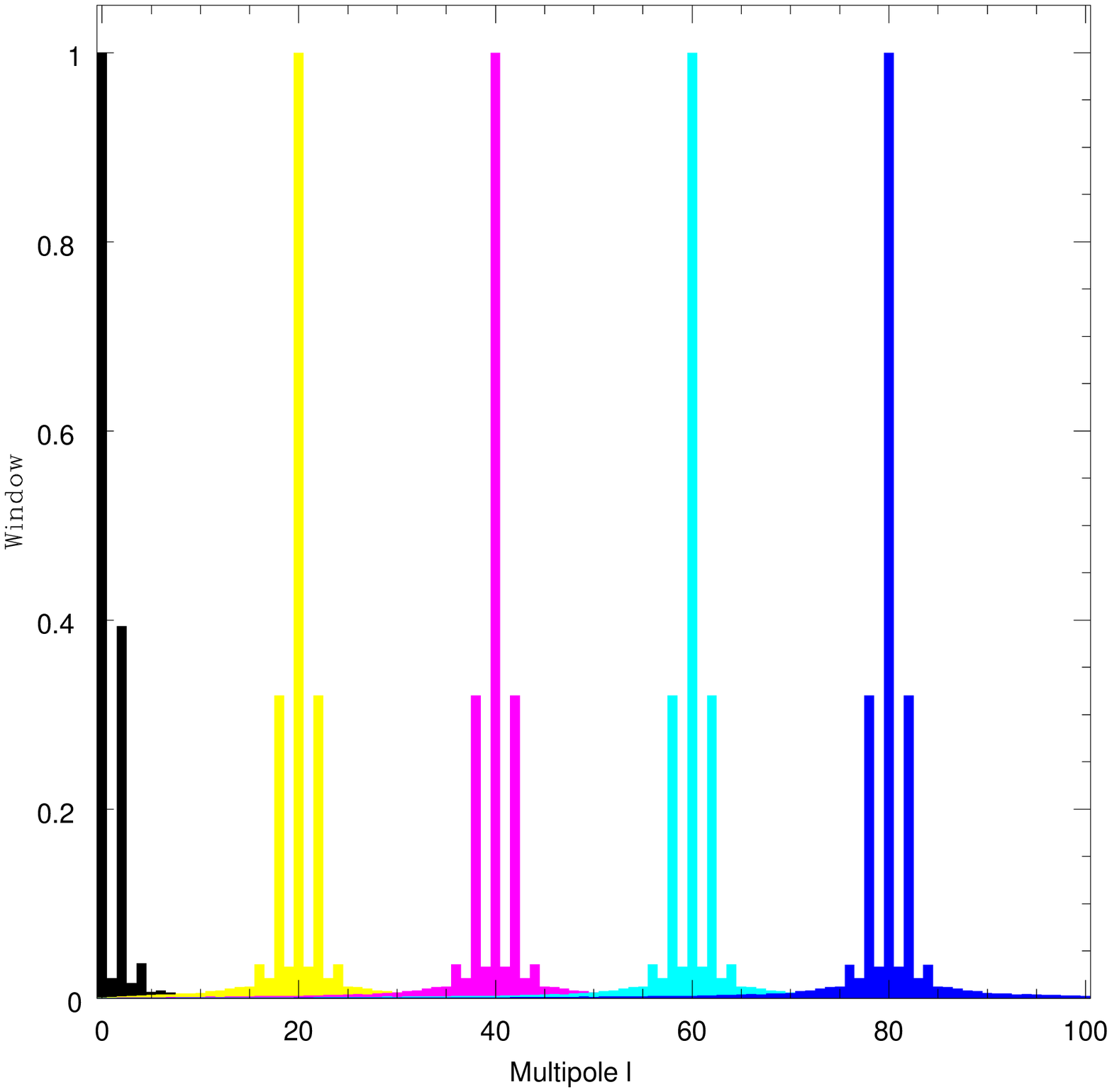}}
\caption[1]{\label{windowFig}\footnotesize%
Sample band power window functions are shown for the cleanest of the sky regions
from \fig{masksFig}, all normalized so that the maximum value is unity.
The approximate lack of leakage from odd numbers of multipoles away
results from the approximate parity symmetry of this region.
}
\end{figure}

\subsection{Blind analysis and the power spectrum matrix}

We compute the power spectrum matrix $\C_l$ in practice 
using the method of \cite{Hivon}; a 
similar approach was used by the Boomerang 
\cite{Ruhl02} and WMAP 
\cite{hinshawpower03} teams.
This simply consists of expanding the masked sky patch in question 
in spherical harmonics and then correcting for window function effects.
Our only variation is that we do not invert the window matrix to obtain
anticorrelated band power estimates with delta function window functions.
Rather, we simply divide by the quadratic estimators by the area of the
window $\delta\T_\l^2$ window functions, which asymptotes to 
the unbiased minimum variance estimators of \cite{cl} on scales much smaller
than the sky patch analyzed. An example of our window functions is shown 
in \fig{windowFig}.

As an example,
\fig{channel4powerFig} 
shows the measured power spectrum for
the
V-band, the cleanest of WMAP's frequency bands.

One fact worth emphasizing is that our weighting scheme of \eq{OptimalwEq} is 
totally blind, assuming nothing whatsoever about the
CMB power spectrum, 
the foregrounds, the WMAP detector noise or external templates.
The only assumption is that the CMB spectrum is the blackbody that the WMAP team have modeled it as,
so that the CMB contributes equally to all five 
channels --- otherwise the vector $\e$ would be replaced by some other constant vector.
We see that there is no need to model the CMB, the foregrounds or the noise, since all we need for computing
the optimal weights is the {\it total} power spectrum matrix $\C_l$, containing the total contribution
from CMB, foregrounds and noise combined --- and this can be measured directly from the data.

We can decompose $\C_l$ as a sum of two terms,
\beq{CdecompEq}
\C_\l = \Cljunk + \CClcmb = \Cljunk + \Clcmb\e\e^t,
\eeq
where the second term is the CMB contribution and the first term is the contribution from 
noise and foregrounds.
Note that if we keep $\Cljunk$ fixed and change $\Clcmb$, 
the weights given by \eq{OptimalwEq} stay the same.
The easiest way to see this is to note that 
the quantity we are minimizing is 
$\expec{|\alm|^2}=\w^t\Cl\w = \w^t\Cljunk\w+\Clcmb(\e\cdot\w)^2 = \w^t\Cljunk\w+\Clcmb$,
so the CMB power is just an
additive constant that does not affect the optimal weighting.
This means that
our method is
blind to assumptions about the 
underlying (ensemble-averaged) CMB power spectrum.
Although we will return below in \Sec{InterpretationSec} to the issue of how 
to determine what fraction of the power $\C_\l$ comes from 
each of the two terms in \eq{CdecompEq}, it is important to remember 
that this affects only the
physical interpretation, not our cleaning method and the maps we produce.

\subsection{Subdividing the sky}
\label{MaskSec}

To minimize the variance in our cleaned map, we should take advantage of all ways in which the 
unwanted signals (noise and foregrounds) differ from the CMB in their contribution to the
covariance matrix $\N$ in \eq{ComboEq1}.
Above we exploited their different dependence on angular scale $\l$. 
Unlike the CMB, foregrounds are not an isotropic Gaussian random field.
Rather, their variance differs dramatically between clean and dirty regions of the sky.
It is therefore desirable to subdivide the sky into a set of regions of increasing cleanliness and 
perform the cleaning separately for each one \cite{wiener}. One then expects our method described above to 
settle on more aggressive weights for the dirtier regions, where foregrounds are much more of a concern than noise.
A second advantage of such a subdivision is that the frequency dependence of the foregrounds is likely to differ between 
very dirty and very clean regions, again resulting in different optimal weights. The WMAP team used the latter argument 
to motivate their subdivision of the sky into 12 regions, and convincingly demonstrated that foreground spectra
indeed did vary across the sky, notably for synchrotron radiation where the spectrum was found to 
steepen towards increasing galactic latitudes.

\begin{figure} 
\centerline{\epsfxsize=\figsize\epsffile{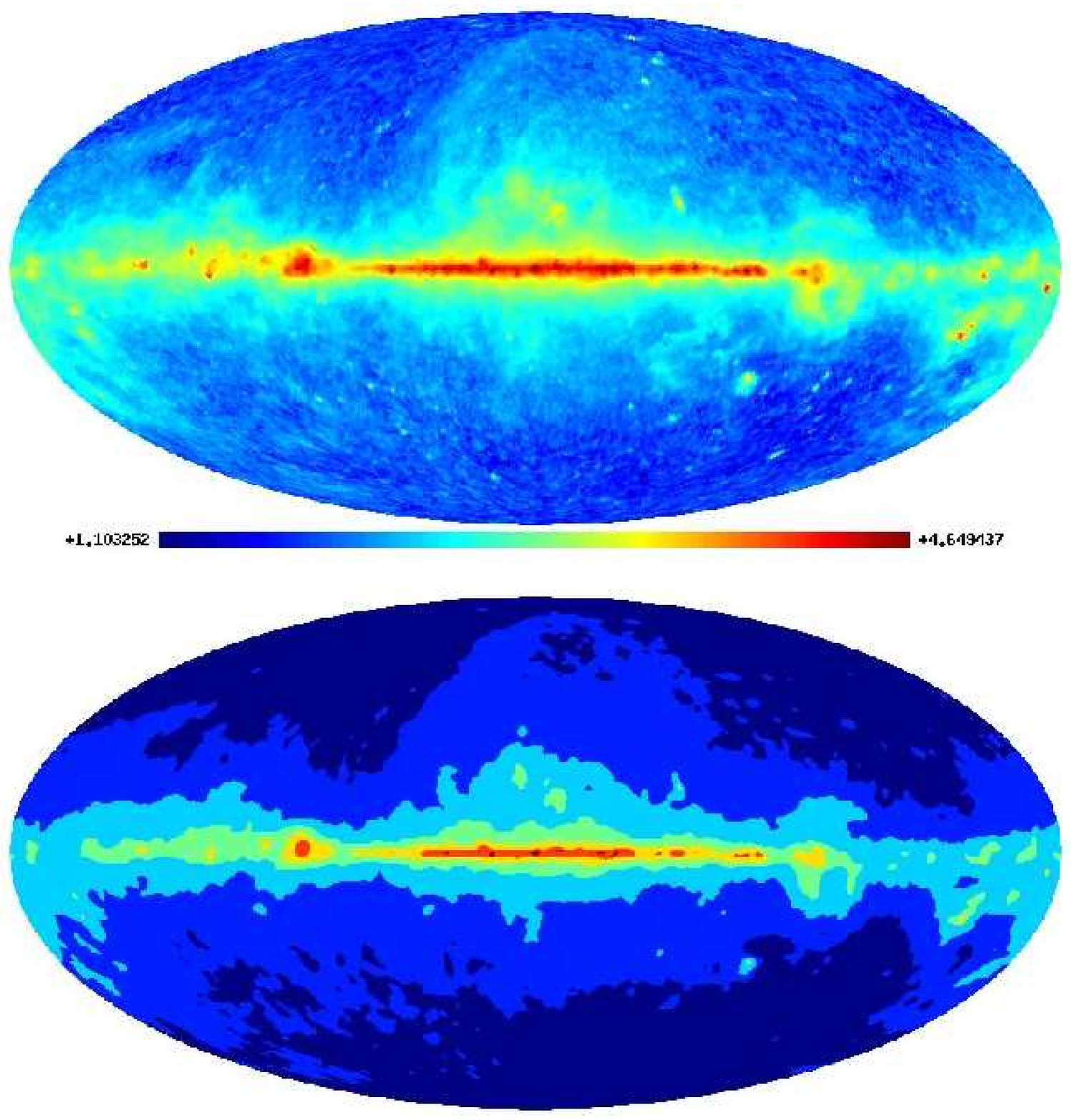}}
\caption[1]{\label{masksFig}\footnotesize%
The top panel shows our junk map with a color scale that 
is uniform in the logarithm of the temperature in $\mu$K.
The bottom panel show our subdivision of the sky into seven regions 
of decreasing cleanliness. From outside in, they correspond to junk map 
temperatures $T<100\uK$, $100\uK-300\uK$, $300\uK-1\mK$, $1\mK-3\mK$, 
$3\mK-10\mK$, $10\mK-30\mK$ and $T>30\mK$, respectively.
The last of these regions contains only the set of roundish
blobs in the inner Galactic plane. 
The second dirtiest region is seen to be topologically 
disconnected, and we treat its leftmost and rightmost
blobs as separate regions, giving nine regions in total.
}
\end{figure}

The WMAP team have created a set of sky-masks of increasing cleanliness based solely the K-band map.
Although these masks are undoubtedly fine in practice, the procedure used in creating them is not blind,
since it rests on the assumption that
all dirty regions are dirty in the K-band.
In particular, if a foreground manifests itself only at higher frequencies 
(imagine, say, a blob with localized dust emission without detectable synchrotron or free-free emission), 
it would go unnoticed. 
A second minor drawback of this K-band approach is that random CMB fluctuations affect the
masks at a low level. In other words, the mask was based on the upper left map in
Figure 2 which, as opposed to the upper right map, contains CMB fluctuations.

To preserve the blind nature of our method, we therefore create sky masks with a different procedure.
We first form four difference maps W-V, V-Q, Q-K and K-Ka, thereby obtaining maps guaranteed to be free of
CMB signal that pick up any signals with a non-CMB spectrum.
We then form a combined ``junk map'' by taking the largest absolute value of these four maps at each point
in the sky (\fig{masksFig}, top). Finally, we create disjoint sky regions based on contour plots of this map.
We use cuts that are roughly equispaced on a logarithmic scale, corresponding to thresholds of
30000, 10000, 3000, 1000, 300 and 100$\mu$K (\fig{masksFig}, bottom).
We emphasize that we have found no evidence whatsoever for any actual problems with the WMAP 
team masks, and opt to use our own simply to preserve the blind and CMB-independent nature of our analysis.
As a cross-check, we also repeated our entire analysis using the WMAP masks Kp0, Lp2 and Kp12, obtaining 
similar results.

We followed the WMAP team procedure in the details of converting the contours into the masks shown
in \fig{masksFig} (bottom): we downsampled the
junk map to HEALPpix resolution 64, imposed the cuts, went back up to HEALPix resolution 512, 
performed Gaussian smoothing with FWHM=$2^\circ$ on the $\{0,1\}$-valued mask and imposed a cutoff of 0.5.
The WMAP team reported strong spatial variations of foreground spectra in the innermost parts
of the galactic plane, and therefore subdivided this into 11 disjoint regions.
We were unable to
reproduce
this procedure since they did not specify which these regions were.
Instead, we
merely lopped off three spatially disconnected islands in the two dirtiest regions 
as their own separate masks, as illustrated in \fig{masksFig} (bottom), leaving 9 separate masks in total.

Our multipole-based cleaning is nonlocal in the spirit of \eq{WdefEq}, and 
although it guarantees that the CMB signal is preserved separately for each pixel, this
is of course not the case for foregrounds.
To avoid
mirages
of foreground emission from the Galactic plane
leaking
up to high latitudes, we clean the galaxy ``from inside out'', \ie, clean the dirtiest region first, 
the second dirtiest region second, {\etc} 
To be specific, we start by defining five temporary maps, 
initially set equal to the five WMAP channels
(Figure 2, left). We then   
repeat the following cleaning procedure nine times, once corresponding to each region $i=1,...,9$:
\begin{enumerate}
\item Compute the power spectrum matrix $\C_\l$ and the optimal weights $\w_\l$ for the $\ith$ region only.
\item Expand the five temporary maps in spherical harmonics and compute a cleaned all-sky map using the weights from step 1.
\item Replace the $\ith$ region of the temporary maps by the corresponding region in the cleaned map from step~2
      (smoothed to the resolution appropriate for that channel, of course, by multiplying
       by $B_\l^i$ in $\l$-space).
\end{enumerate}
Visually, one thus sees the foreground contamination gradually being cleaned off from Figure 2
as this iteration proceeds, starting in the inner Galactic plane and proceeding outward.
At the end of the nine iterations, the five temporary maps equal the desired cleaned map at the
K-, Ka-, Q-, V-, and W-band resolutions, respectively.

\subsection{Interpretation of the cleaned maps}
\label{InterpretationSec}

Figure 1 (bottom) shows
our
final cleaned map and \fig{latitude_comparisonFig}
shows
its power spectrum in sky regions of varying cleanliness. We plot all maps after
21$^\prime$  Gaussian smoothing (giving a net FWHM of
24$^\prime$)
to prevent them from being 
undersampled by the pixels in the image. 
The reader interested in using this map can download the corresponding 13 Megabyte HEALPix fits file
from the web\footnote{Our cleaned maps are available for download at 
\protect\url{http://www.hep.upen.edu/~max/wmap.html},
together with a high-resolution version of
this paper.}.
To use this map, it is important to be clear on how to interpret it.

First of all, it is a sum of CMB, foreground and detector noise fluctuations.
Although it was constructed by minimizing its power 
spectrum $\Clclean=\w_\l^t\C_\l\w_\l$, 
its power nonetheless 
gives a
{\it lower\/}
bound on the CMB power spectrum, since the minimization was performed 
subject to the constraint that the CMB be preserved;
$\e\cdot\w=1$\footnote{
The only way in which its power could be biased low would be if random fluctuations in
our estimate of the $\C_\l$ matrix conspired to remove power. Although we found no indication
of this actually happening, we computed our weights $\w_\l$ using a heavily smoothed version of
$\C_\l$-matrix as a precaution. Specifically, we smooth over at least $\Delta\l=10$ or
100 $(\l,m)$-modes, 
whichever is larger, obtaining an $\l$-dependence of $\C_\l$ with no 
visible trace of random fluctuations.}.

So what fraction of the power seen in \fig{latitude_comparisonFig} 
is due to CMB, foregrounds and detector noise, respectively?
Let us first get some rough estimates from the figures, then present more
quantitative limits.
Since this paper is focused on minimizing foregrounds, not on physically modeling them,
the interested reader is
referred to the detailed foreground study of the WMAP team \cite{BennettForegs} for
more information.

\begin{figure} 
\centerline{\epsfxsize=\figsize\epsffile{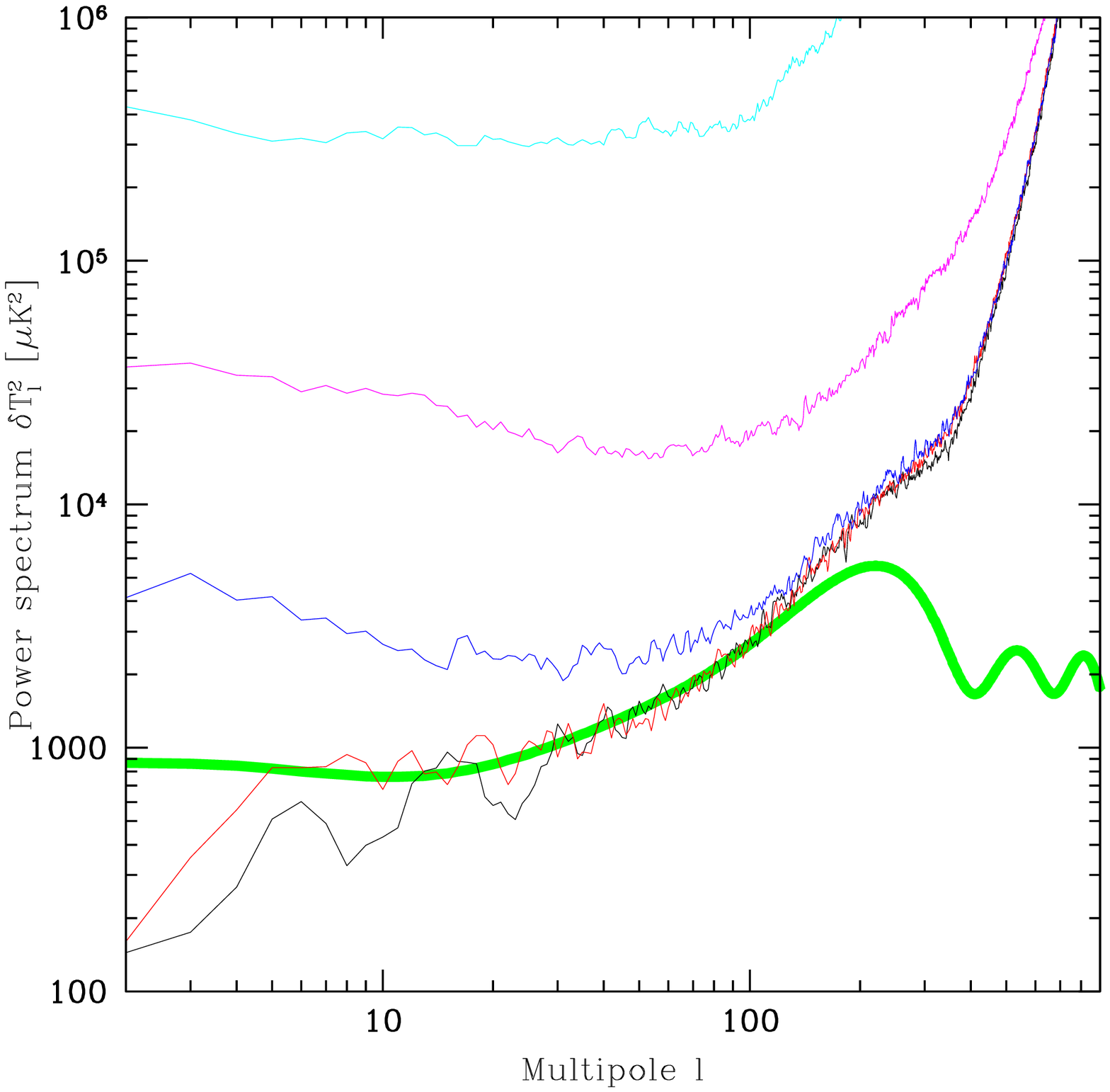}}
\caption[1]{\label{channel4powerFig}\footnotesize%
Total power spectra for CMB, foregrounds and noise combined are shown
for the 61 GHz V-band, the one with the overall lowest foreground levels.
This is the $(4,4)$ element of the power spectrum matrix $\C_\l$.
From bottom to top, they correspond to the five cleanest sky regions shown in
\fig{masksFig}. For comparison, the thick curve shows the best-fit CMB model
from \cite{Spergel03,verde03}.
}
\end{figure}

\begin{figure} 
\centerline{\epsfxsize=\figsize\epsffile{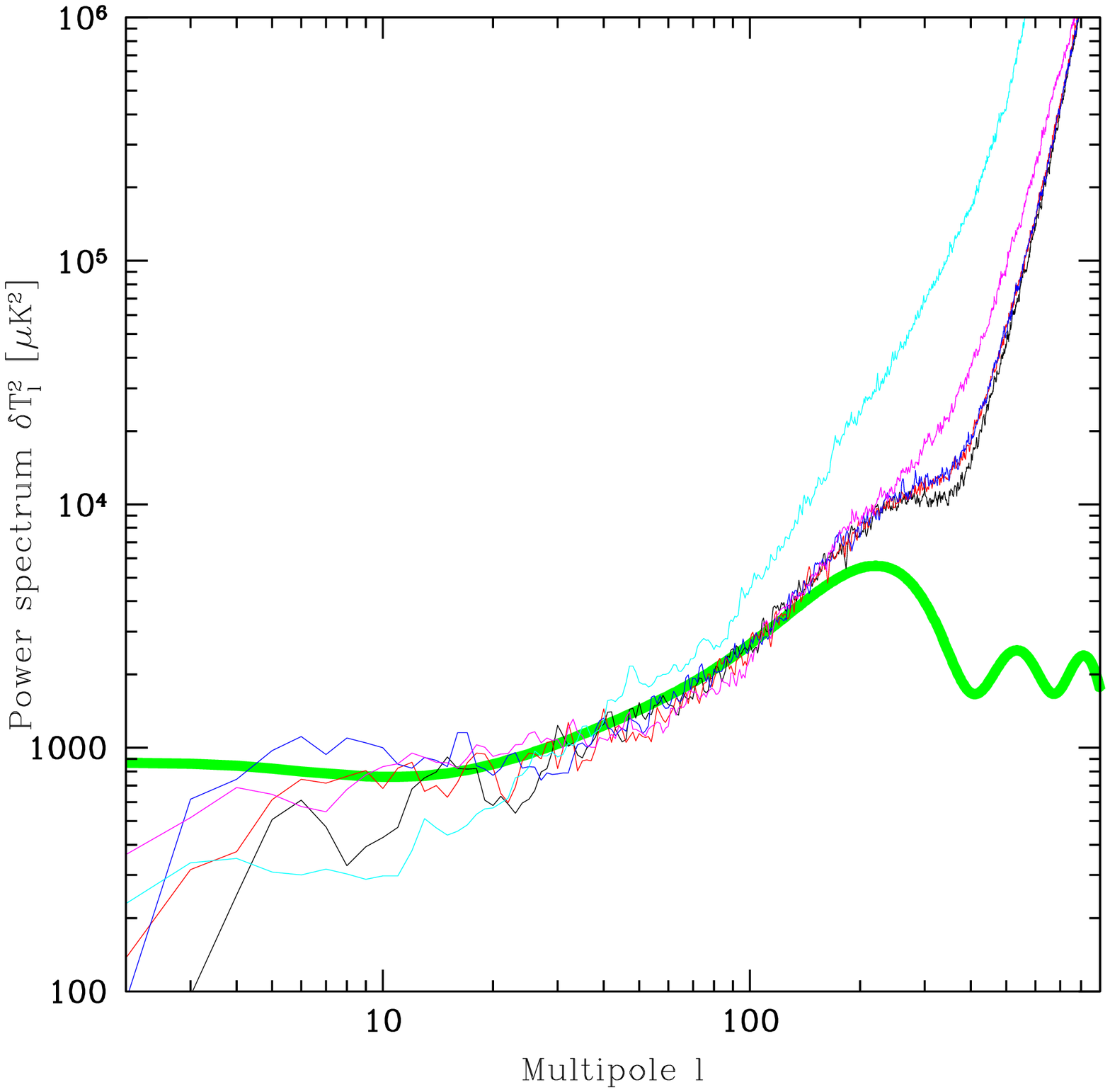}}
\caption[1]{\label{latitude_comparisonFig}\footnotesize%
Same as previous figure, but for our foreground-cleaned CMB map (bottom panel of Figure 1).
The power spectra of the cleanest sky regions are seen to be virtually identical
with those for V-band on large scales, showing how subdominant smooth foregrounds
are at 61 GHz.
}
\end{figure}

\begin{figure} 
\centerline{\epsfxsize=\figsize\epsffile{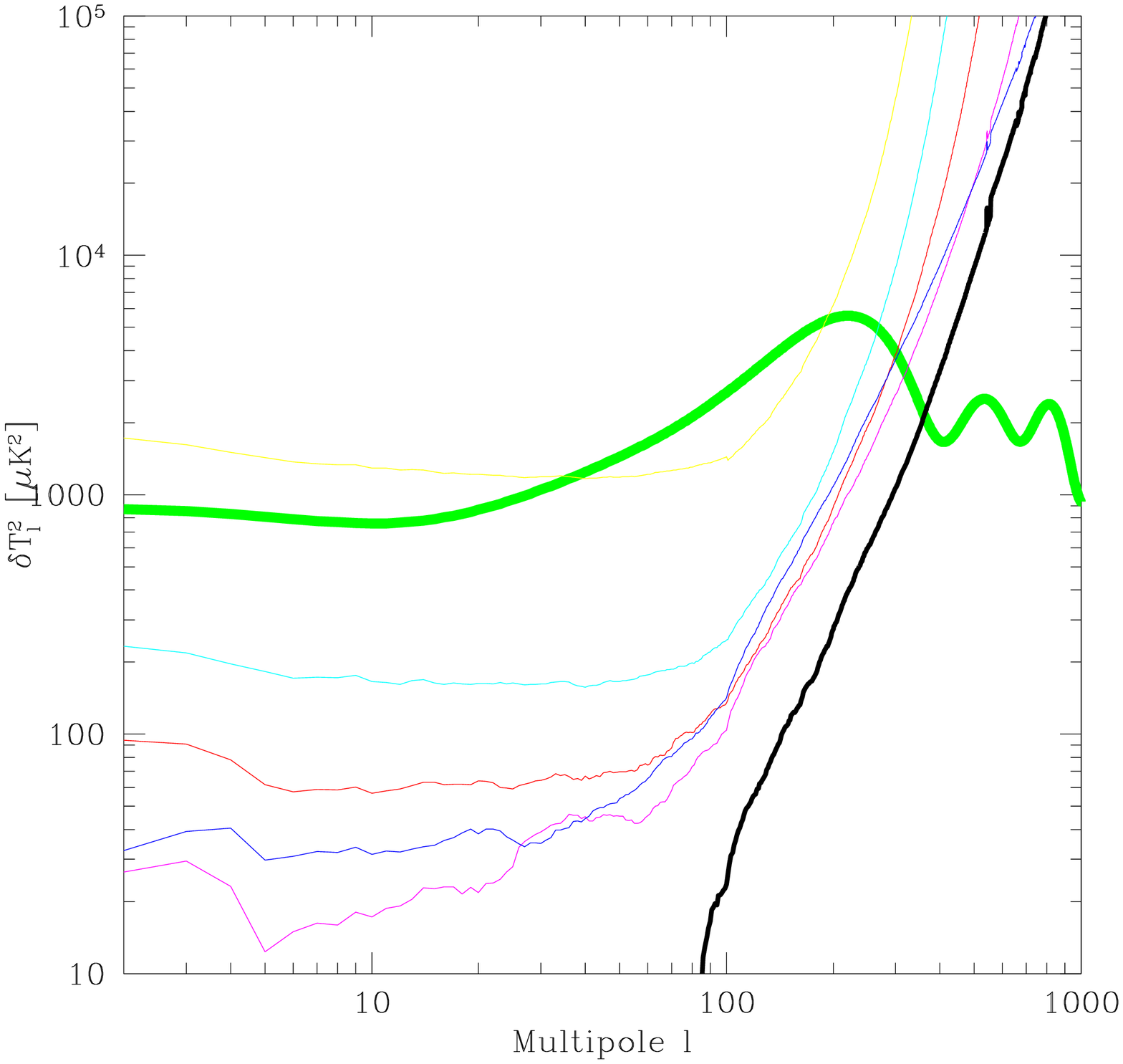}}
\caption[1]{\label{foregpowerFig}\footnotesize%
Power spectra of non-CMB signals (foregrounds plus detector noise) in the cleanest part of the sky shown
in \fig{masksFig}.
The five thin curves show the power spectra of five maps on the right side of Figure 2,
\ie, the five WMAP channels 
after our cleaned CMB map (Figure 1, bottom) has
been subtracted out. From top to bottom, the five curves correspond to 
22.8, 33.0, 40.7, 93.5 and 60.8 GHz, respectively (note that the 
second highest frequency, V-band, is the cleanest).
These curves should be interpreted as lower limits on foregrounds plus noise.
The black curve gives the lower limit on foregrounds plus noise in our cleaned map
using the method described in the text. 
}
\end{figure}

The WMAP team cleverly eliminated the average noise contribution by using only cross-correlations between
different differencing assemblies (DAs) to measure the power spectrum, and used a combination
of sky cutting and foreground subtraction (with external templates) to minimize
the foreground contribution. The best fit cosmological 
model \cite{Spergel03,verde03} to their 
measured CMB power spectrum \cite{hinshawpower03} is shown for comparison in 
figures~\ref{channel4powerFig} and~\ref{latitude_comparisonFig}. 
For $\l\simlt 100$, it is seen to agree well with the lower envelope of 
our curves in \fig{channel4powerFig}, 
suggesting that foregrounds are subdominant in the
cleanest parts of the 61 GHz sky. \Fig{latitude_comparisonFig} shows no noticeable 
excess
power at $\l \l\simlt 100$
due to foregrounds in the cleaned map in any of
the
four cleanest sky regions
from \fig{masksFig}, which together cover all but the very innermost Galactic plane.

The slight power deficit on the very largest scales has two causes: one is the
low quadrupole,
to which we return in \Sec{QuadrupoleSec} below, which pulls down 
neighboring band power estimates
because the band-power window, illustrated in Figure~\ref{windowFig},
contains small off-diagonal contributions.
The second cause is that we have not corrected for the effects 
of monopole and dipole removal,
which pulls down the power estimates on the scale of the sky patch in question 
(our method produces an unbiased CMB map regardless
of
what $\C_\l$ we use in \eq{OptimalwEq},
so this merely raises the variance in the cleaned map above the optimal level).
On smaller scales, detector noise starts to dominate, and is seen to push our 
curves way above the CMB curve.
A worthwhile future project for further quantifying foregrounds would be to repeat 
our analysis with $\C_\l$ estimated in a way that removes the detector noise contribution.
For Q, V and W bands, which each have more than one DA, this can be done using cross-correlations.
For the $(1,1)$ and $(2,2)$ components of $\C_\l$ (the K and Ka power spectra), it would involve
subtracting the noise power using the WMAP team's noise model.

We can, however, give some quantitative limits even based on our measured $\C_\l$ alone.
Grouping the five coefficients $\alm^i$ into a five-dimensional vector $\a_{\l m}$, 
\eq{CmatrixEq} becomes simply $\C_\l = \expec{\a_{\l m}^* \a_{\l m}^t}$, and the
cleaned procedure can be written $\alm = \w\cdot \a_{\l m}$.
Using \eq{OptimalwEq} shows that the power in the cleaned map is 
\beq{ClcleanEq}
\Clclean = {1\over \e^t\C_\l^{-1}\e}.
\eeq
By subtracting our cleaned map from the input map, we obtain maps showing in Figure 2 (right).
These are guaranteed to be free from CMB power, since the cleaned map gave a lower limit on 
the CMB. Most likely, we have subtracted some foreground power too, so the five
maps should be interpreted as placing lower limits on the foreground power.
(As a toy example, imagine synchrotron,
free-free
and dust emission tracing each other
perfectly; the sum of their three spectra can then be written as a constant, which
our method will interpret as CMB, plus a non-negative residual, which our method will interpret 
as foregrounds).
Using \eq{CdecompEq}, the covariance matrix of these five
CMB-free
``junk maps'' is
\beq{CjunkEq}
\Cljunk \ge \Clnocmb\equiv \C_\l - \Clclean\e\e^t = \PP_\l\C_\l\PP_\l^t,
\eeq
where the projection matrix 
\beq{ProjectionDefEq}
\PP_\l\equiv \I-\e\w_\l^t
\eeq
satisfies $\PP_\l^2=\PP_\l$, $\PP_\l\e=0$ and $\PP_\l^t\w=0$ and can be interpreted as projecting out
the CMB component. In this notation, the maps in Figure 2 (right) are defined by 
simply $\PP_\l\a_{\l m}$. The inequality $\Cljunk \ge \Clnocmb$ refers
only
to the diagonal
elements of these matrices.
The five diagonal elements of $\Cljunk$ are plotted in \fig{foregpowerFig} for our cleanest sky region 
from \fig{masksFig}.

How much of $\C_\l$ can possibly be due to CMB? In other words, forgetting for a moment 
about
the
weighting that gave \eq{ClcleanEq}, how large can we make $\Clcmb$ in 
\eq{CjunkEq} before the covariance matrix $\Cljunk$ gets unphysical properties?
First of all, no covariance matrix can have negative eigenvalues, so we we must stop
increasing $\Clcmb$ once the smallest eigenvalue of $\Cljunk$ drops to zero.
In fact, $\PP^t\w=0$ implies that when using our optimal weighting and hence \eq{ClcleanEq},
$\Cljunk$ has a vanishing eigenvalue corresponding to the vector $\w$, and it is easy to
show that this alternative method for estimating $\Clcmb$ is equivalent to 
our original method, being simply an alternative derivation of \eq{ClcleanEq}.

Let us now make a second assumption: that $\Cljunk$ cannot have any negative 
elements.
The noise covariance matrix is guaranteed to have this property, since the absence of
correlations between bands implies that it is diagonal.
The foreground covariance matrix is also guaranteed to have this property if foreground emission
is indeed {\it emission}, \ie, if foregrounds can make only positive contributions 
to the sky maps. Pure absorption at all frequencies likewise give only positive correlations.
This assumption, made also in the Maximum-Entropy analysis of \cite{BennettForegs}
is likely to be valid WMAP frequencies, since the
only known exception is the thermal SZ effect, and it changes sign only outside of
the WMAP frequency range (around 217 GHz).
In summary, we therefore have two separate limits on the CMB power spectrum:
\beqa{CcmbLimEq}
\Clcmb &\le& \Clclean,\\
\Clcmb &\le& \min_{i,j}(\C_\l)_{ij}\label{CcmbLimEq2}.
\eeqa
$\Clclean$ is also the actual variance of the cleaned map, 
so if the second limit is lower than first, then we know that the difference 
cannot be due to CMB. 

To gain intuition, consider the following two examples
of the junk + CMB decomposition of \eq{CdecompEq}
for the simple case of only two frequency bands:
\beqa{DecompositionExampleEq}
\C_\l =
\left(\bs\begin{tabular}{cc}
5&3\\
3&2
\end{tabular}\bs\right)
&=&
\left(\bs\begin{tabular}{cc}
4&2\\
2&1
\end{tabular}\bs\right)
+
\left(\bs\begin{tabular}{cc}
1&1\\
1&1
\end{tabular}\bs\right)\\
\C_\l =
\left(\bs\begin{tabular}{cc}
2&1\\
1&2
\end{tabular}\bs\right)
&=&
{1\over 2}\left(\bs\begin{tabular}{cc}
1&-1\\
-1&1
\end{tabular}\bs\right)
+
{3\over 2}\left(\bs\begin{tabular}{cc}
1&1\\
1&1
\end{tabular}\bs\right)\label{sillyDecompEq}\\
&=&
\left(\bs\begin{tabular}{cc}
1&0\\
0&1
\end{tabular}\bs\right)
+
\left(\bs\begin{tabular}{cc}
1&1\\
1&1
\end{tabular}\bs\right)
\eeqa
In the first case, \eq{ClcleanEq} gives 
$\Clcmb=1$, \ie, the largest contribution we can possible 
attribute to the CMB (the second term) is $\e\e^t$ --- if we 
scaled up the last term, then $\Cljunk$ (the first term)
would acquire an unphysical negative eigenvalue.
Moreover, we see that $\Cljunk$ looks like the covariance 
matrix of a perfectly respectable foreground, 
with a spectrum such that it contributes twice as high rms fluctuations
to the first channel as to the second, and 
with perfect correlation between the two (the dimensionless correlation
coefficient is $r=1$).
Such a foreground would be completely removed by our method, and
so we cannot place any lower limit on the foreground contribution
to the cleaned map in this case.
In the second case, \eq{ClcleanEq} gives 
$\Clcmb=3/2$, so this will be the variance in the cleaned map.
However, $\Cljunk$ (the first term) in \eq{sillyDecompEq}
exhibits an unphysical anticorrelation between the two bands which
neither noise nor foregrounds could produce, and
\eq{CcmbLimEq2} tells us that the CMB power $\Clcmb\le 1$,
corresponding to the alternative decomposition on the last line.
This means that the junk map must have a power contribution of
at least $1/2$ which is not due to CMB.

This difference, which places a lower limit on the amount of
non-CMB signal in our cleaned map, is shown as the 
black line in \fig{foregpowerFig} for our actual five-band case.
We see that although we get an interesting lower
limit
for $\l>100$, presumably dominated by 
detector noise, the residual foreground level 
is consistent with zero at low $\l$.

\subsection{Wiener filtering}

Figure 1 (bottom) shows our cleaned map of the CMB sky, and this is the map
that should be used for cross-correlation analysis with other data sets and other
scientific applications.
For visualization purposes, however, we can do better. 
The ``best-guess'' map of what the CMB looks like, in the sense of minimizing the
rms errors, is the Wiener filtered map defined by \cite{Wiener49}
\beq{WienerEq}
\alm^{\rm Wiener} = {\Clcmb\over\Clclean}\alm.
\eeq
This staple signal processing technique, multiplying by ``signal over signal plus noise'',
has additional attractive properties; for instance, it constitutes the best-guess
(maximum posterior probability)
map in the approximation of Gaussian fluctuations.
Examples of recent applications of Wiener filtering to CMB and galaxy mapping include 
\cite{Rybicki92,Lahav94,Fisher95,Zaroubi95,Bond95,Bunn96,saskmap}.
Our resulting Wiener filtered map is shown in Figure 1 (middle),
using the best fit model from the WMAP team \cite{Spergel03,verde03} as our estimate
of $\Clcmb$ in the numerator of \eq{WienerEq}.
For the denominator, we take
the larger of our measured
$\Clclean$ and $\Clcmb$,
so that the ratio is guaranteed to be $\le 1$.
The result is 
not an unbiased CMB map. Rather, \eq{WienerEq} shows that each multipole gets multiplied
by a number between 0 and 1, so features with high signal-to-noise are left unaffected
whereas features that are not statistically significant become suppressed.
This means that the features that you see in Figure 1 (middle) are likely to be
real CMB fluctuations, having signal-to-noise exceeding unity.

The cleaned map at the bottom of Figure 1 reveals some residual galactic fluctuations
on very small angular scales, caused mainly by the fact that no other channels have fine enough
angular resolution to help clean the W-band map for very large $\l$. 
We Wiener filter each of the nine 
sky regions from \fig{masksFig} separately, so $\Clclean$ is much higher near the Galactic plane.
This is why the Galactic contamination is imperceptible in the Wiener filtered map:
\eq{WienerEq} automatically suppresses fluctuations in regions with large residual
foregrounds.

\begin{figure} 
\centerline{\epsfxsize=\figsize\epsffile{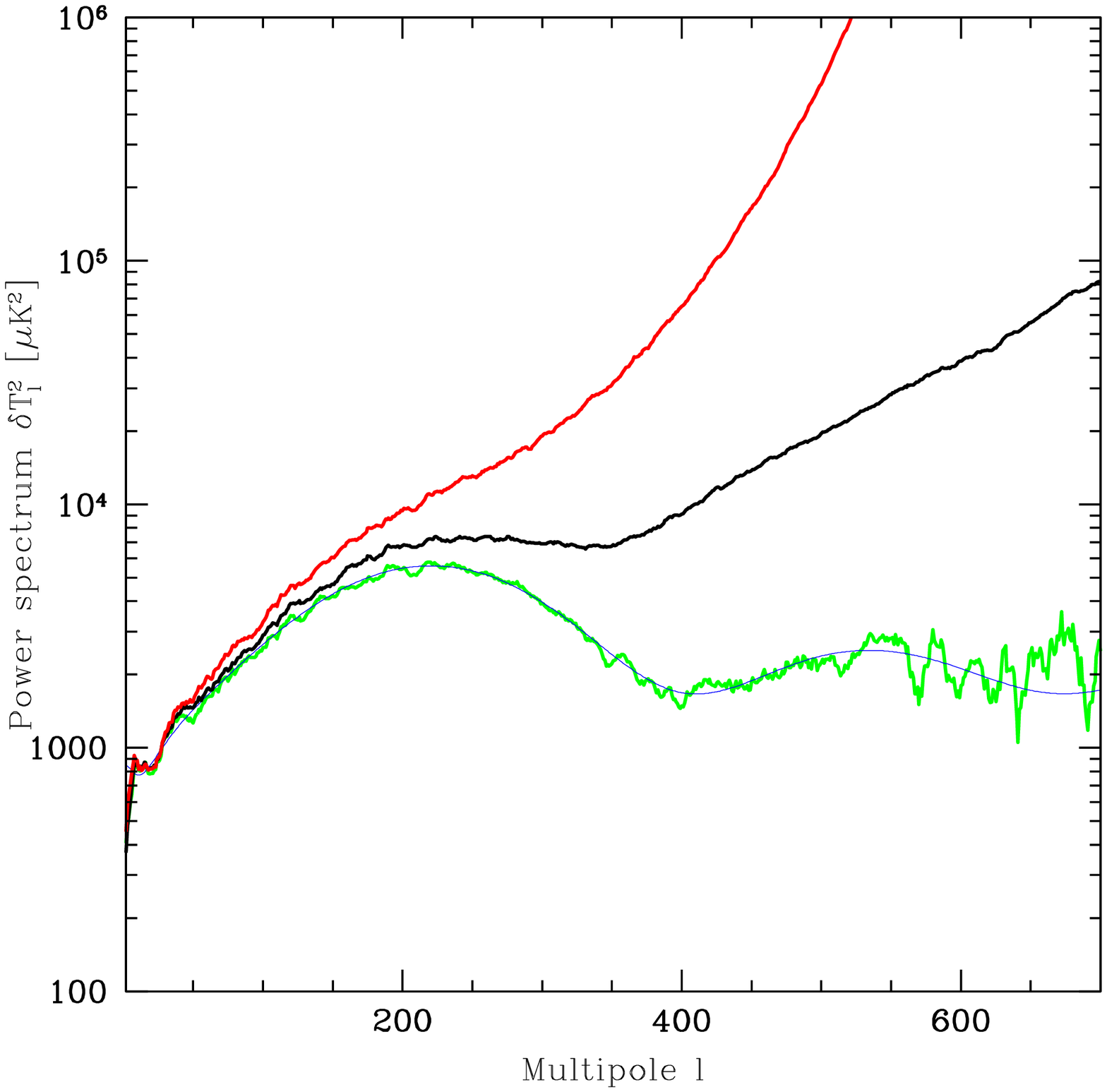}}
\caption[1]{\label{bennett_comparisonFig}\footnotesize%
Comparison of the total (CMB+foregrounds+noise) power spectra of the 
WMAP team internal linear combination map \protect\cite{BennettForegs}
(top curve) and our cleaned map (middle curve), both for our cleanest 
sky region. 
Both of these cleaned maps are seen to reproduce on large scales
the CMB power spectrum measured by the WMAP team 
\protect\cite{hinshawpower03} (lower wiggly curve), which has no net noise contribution
because it is based on cross-correlations between channels.
The lower smooth curve is the WMAP team's best fit model \cite{Spergel03,verde03}.
As explained in the text, 
the noise contribution is seen to become important earlier in the 
WMAP team's cleaned map than in ours because it is limited
by the lowest resolution frequency bands.
All power spectra have been smoothed with a boxchar filter of width 
$\Delta\l=10$ to reduce scatter.
}
\end{figure}

\section{Discussion}
\label{DiscussionSec}
\label{QuadrupoleSec}

We have performed an independent foreground analysis of the WMAP maps to produce a cleaned CMB map.
The only assumption underlying our method is that the CMB contributes equally to all 
five channels. This assumption rests on very solid ground \cite{BennettWMAP}.
The basic reason for this is that the COBE/FIRAS determination of 
the CMB spectrum \cite{Mather94} is based on the absolute CMB signal, which is 
about $10^5$ times larger than the fluctuations that we have considered in this paper.

Figure 1 shows that our map agrees very well with the internal linear combination (ILC)
map from the WMAP team on the scales $\simgt 1^\circ$ where they can be compared.
This is yet another testimony to the high signal-to-noise in the WMAP data and to the fact that
unpolarized CMB foregrounds are manageable: the basic spatial features of the CMB are 
insensitive to the details of the foreground removal method used.

The basic advantage of our map is illustrated in \fig{bennett_comparisonFig}, which compares its
all-sky power spectrum with that of the WMAP team ILC map.
Both power spectra have had beam effects removed here; in Figure 1, our map
is shown at the W-band resolution and the ILC map is shown at $1^\circ$ resolution as released.

The first thing to note from \fig{bennett_comparisonFig} is that foregrounds appear 
highly subdominant in both maps, since their power spectra essentially coincide with
the WMAP CMB power spectrum on large scales where noise becomes unimportant.
Second, as expected, the main improvement in our map is seen to be on the
smaller scales where noise is important, gaining a factor of 30\% at
the first acoustic peak and about a factor of two at the second peak where
noise from the low frequency channels is beginning to exponentially dominate
the ILC map.  

We hope that our map
will
prove useful for a variety of scientific applications.
For cross-correlation with external maps, its lowered noise power should be
particularly advantageous for pulling out small-scale signals, for instance those
associated with lensing and SZ clusters. Rather that attempting detailed modeling
of the residual noise and foreground fluctuations in our map, a simple way to place
error bars on such correlations will be repeating the analysis with a suite of rotated and
flipped versions of our map as in, \eg, \cite{19ghz}.

\begin{figure} 
\centerline{\epsfxsize=\figsize\epsffile{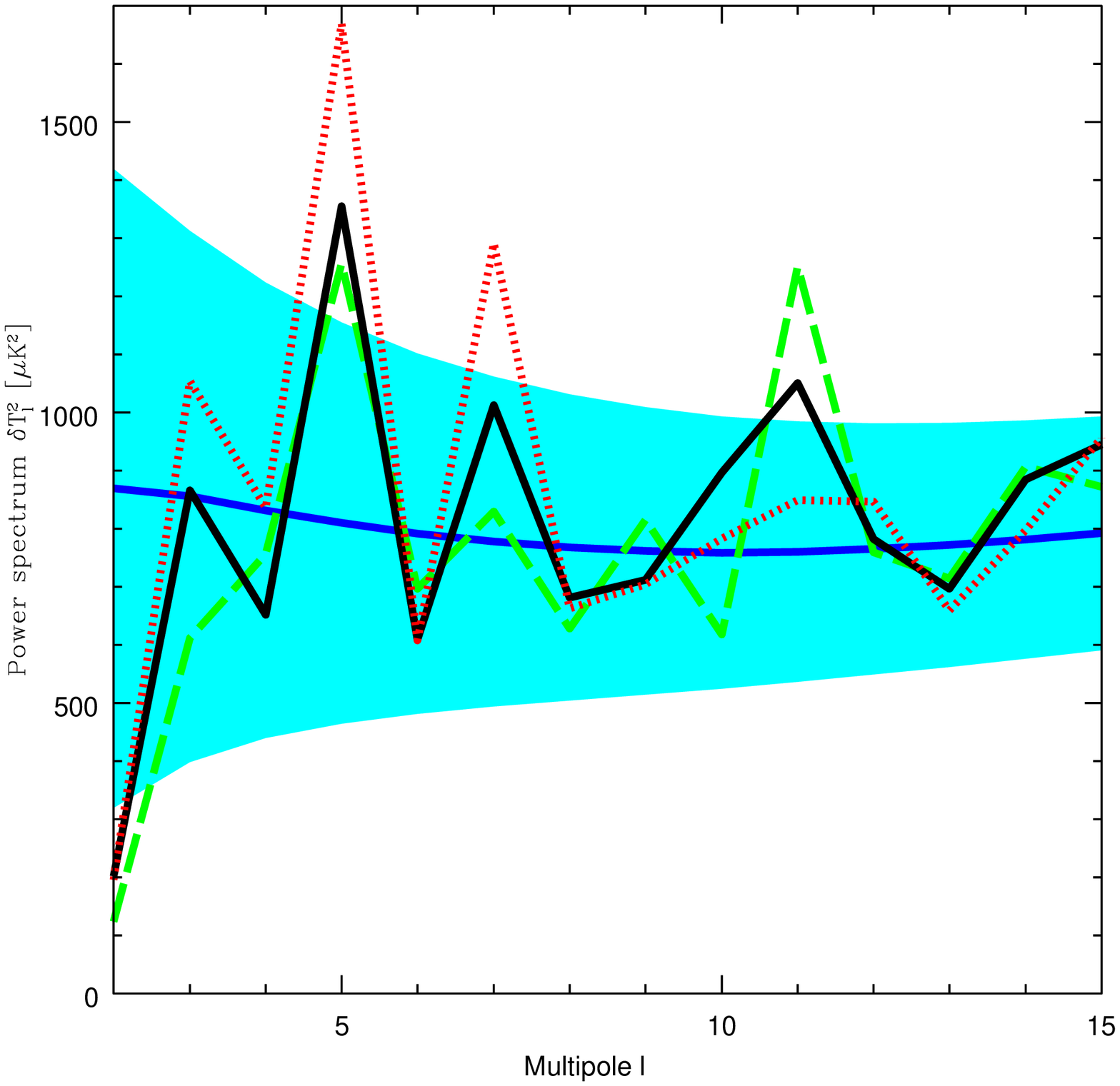}}
\caption[1]{\label{quadrupoleFig}\footnotesize%
A heretic all-sky analysis of our cleaned map (solid jagged curve)
gives a slightly less low quadrupole 
$\delta T_2^2$ than the cut-sky WMAP analysis \protect\cite{hinshawpower03}
(dashed curve), and also agrees well with the quadrupole from 
an all-sky analysis of the WMAP team cleaned map of \protect\cite{BennettForegs} (dotted curve).
The smooth curve shows the WMAP team best fit model
\protect\cite{Spergel03,verde03} with the band indicating the cosmic variance errors
(WMAP noise and beam effects are completely negligible on these scales).
}
\end{figure}

\begin{figure} 
\centerline{\epsfxsize=\figsize\epsffile{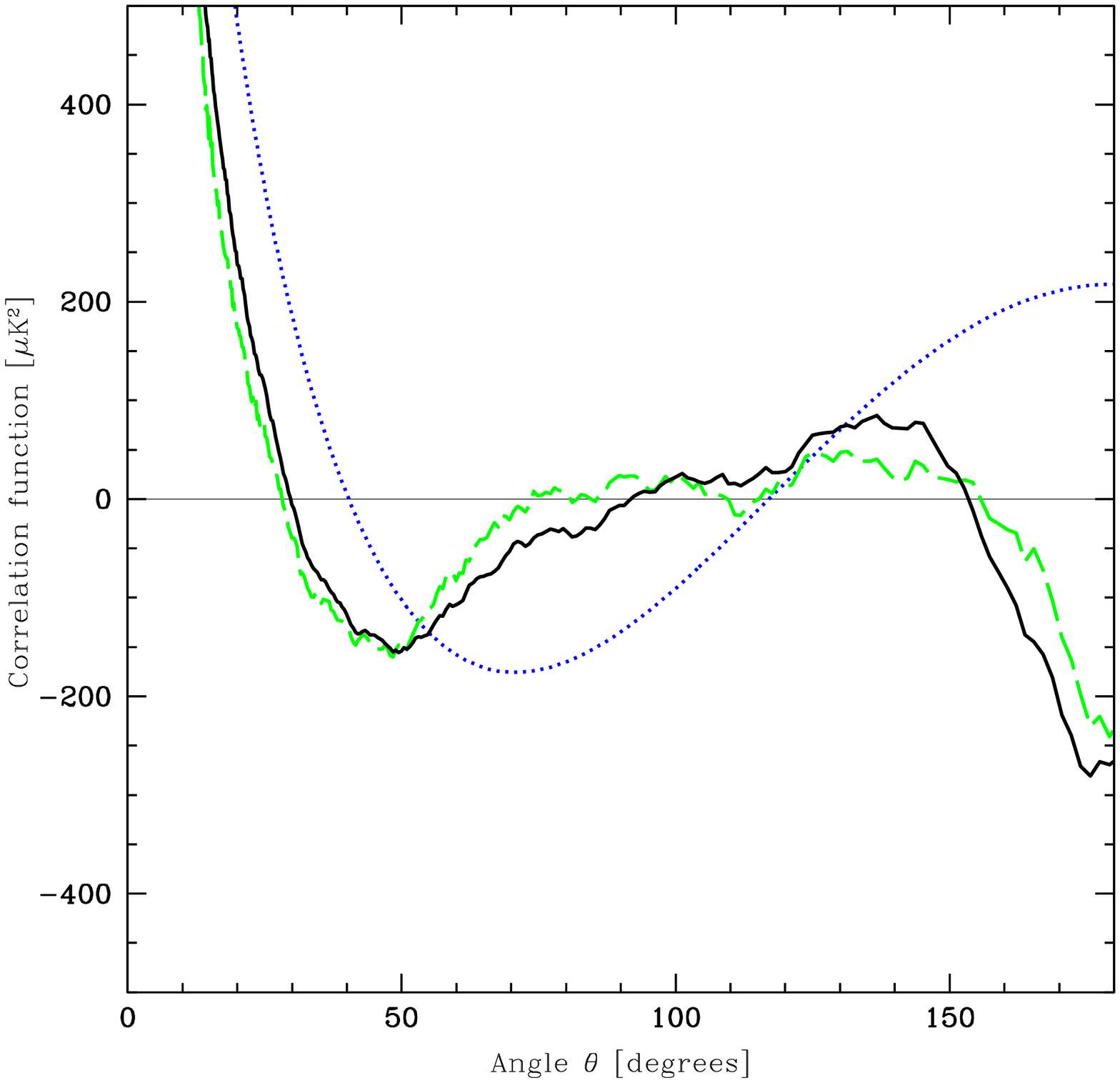}}
\caption[1]{\label{corrfuncFig}\footnotesize%
The angular correlation functions are shown corresponding to 
our all-sky cleaned map (solid curve), 
the cut-sky WMAP power spectrum (dashed curve) 
and the best-fit WMAP team cosmological model
(dotted curve). 
This is simply the Legendre transform of the power spectra from the previous figure,
\ie,
$(4\pi)^{-1}\sum_\l (2\l+1) P_\l(\cos\theta) C_\l$, so 
the the quadrupole and octopole contributions are
shaped as $3\cos^2\theta-1$ and 
$5\cos^3\theta-3\cos x$, respectively.
}
\end{figure}

\begin{figure*} 
\hglue0cm\epsfxsize=17.5cm\epsffile{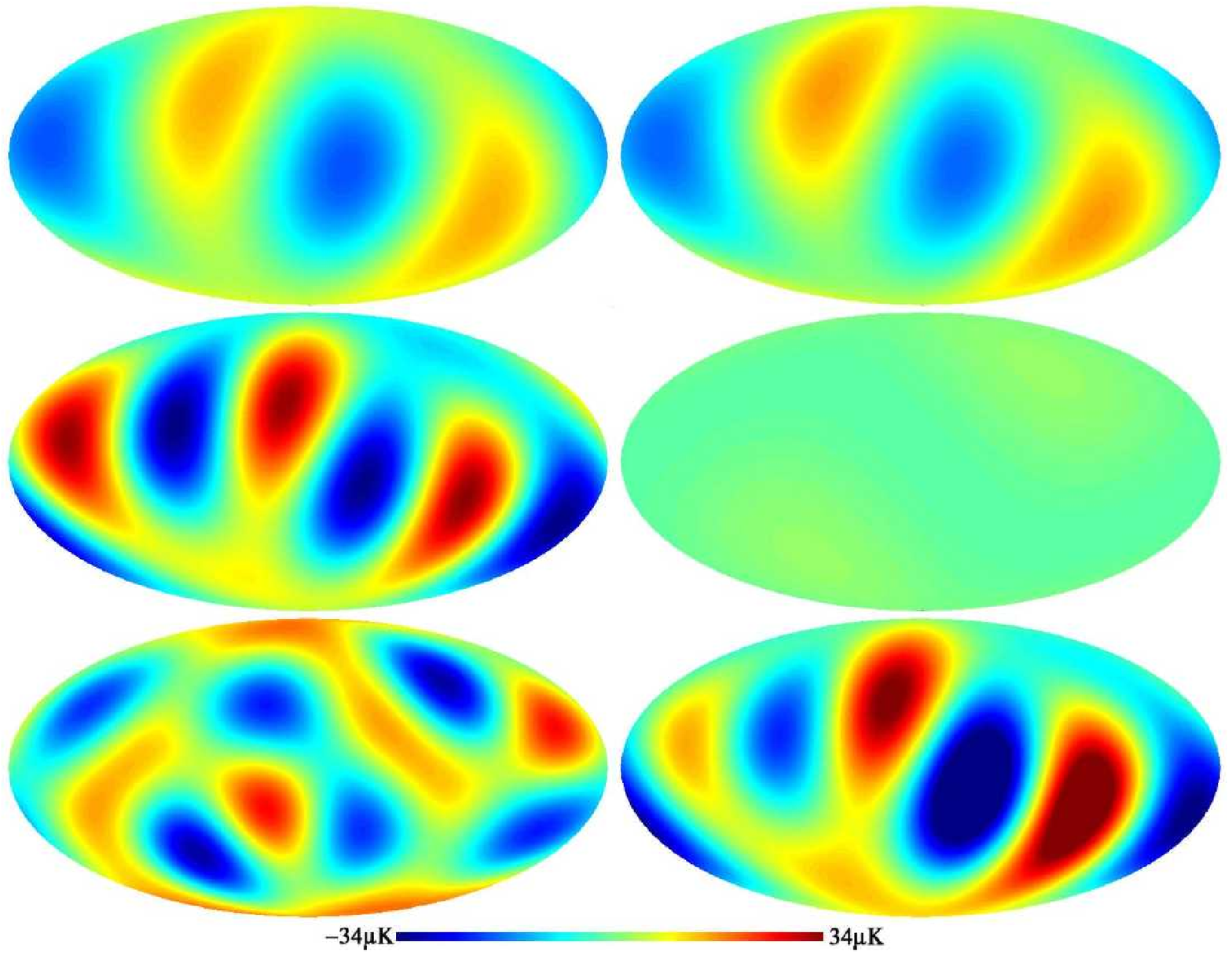}
\smallskip
\caption[1]{\label{multipolesFig}\footnotesize%
The left panel shows the quadrupole (top), octopole (middle) and hexadecapole (bottom) components
of our cleaned all-sky CMB map from Figure 1 on a common temperature scale.
Note that not only is the quadrupole power low, but both it and the octopole have almost all their power
perpendicular to a common axis in space, as if some process has suppressed 
large-scale power in the direction of this axis.
We computed the corresponding images for the WMAP team ILC map as well, and found
them to be very similar.
The right panel shows he cosmic quadrupole (top) after a correcting for a crude 
estimate of the dynamic quadrupole (middle) from our motion relative to the CMB rest frame.
The bottom right map shows the sum of the quadrupole and octopole maps from the left panel.
}
\end{figure*}


\bigskip
\noindent
{\footnotesize
{\bf Table 1} -- 
Measurements of the CMB quadrupole and octopole.
\smallskip
\begin{center}
{\footnotesize
\begin{tabular}{|l|r|r|r|}
\hline
Measurement		&$\delta T_2^2$ $[\mu$K$^2]$	&p-value	&$\delta T_3^2$ $[\mu$K$^2]$\\
\hline		
Spergel {\etal} model	&869.7  	        	&		& 855.6\\
Hinshaw {\etal} cut sky &123.4	     	        	&1.8\%		& 611.8\\
ILC map all sky 	&195.1	     	        	&4.8\%		&1053.4\\
Cleaned map all sky	&201.6	     	        	&5.1\%		& 866.1\\
Cosmic quadrupole	&194.2	     	        	&4.7\%		&\\
Dynamic quadrupole	&  3.6	     	        	&		&\\
\hline		
\end{tabular}
}
\end{center}
}

Let us close by returning from small to large angular scales.
The surprisingly small CMB quadrupole has intrigued the cosmology community ever since
if was first observed by COBE/DMR \cite{Smoot92}, and simulations by \cite{Spergel03,verde03}
show that the low value observed by WMAP is sufficiently unlikely to warrant serious concern.
The WMAP team measured the quadrupole using only the part of the sky outside of their Galactic cut,
and stress that the dominant uncertainty in its value is foreground modeling.
While we agree with this assessment, it should be borne in mind that noise variance and beam
issues are completely negligible on these huge angular scales, so this does not imply that
the foreground uncertainties are large compared to the signal itself. Indeed, 
\fig{bennett_comparisonFig} suggests that foregrounds are subdominant to the intrinsic
CMB signal even {\it without any Galaxy cut} if a foreground-cleaned map is used.
While the reader may feel disturbed by the clearly visible Galactic residuals in 
Figure 1 (in both the top and bottom maps), it is important to bear in mind that
these signals are present in only a tiny fraction of the total sky area and therefore 
contribute little to the total power spectrum.

To quantify the impact of the Galactic plane, 
we computed a sequence of all-sky power spectra for both our map and the ILC
map.
We first used the unadulterated maps, then repeated the calculation after repacing 
the dirtiest of the nine regions from
\fig{masksFig} with zeroes, then zeroed our the 2nd dirtiest region as well, {\etc}
This zeroing procedure obviously biases the measurements by removing power, by an amount related 
to the zeroed area, but provides a powerful test of how sensitive the results are
to Galactic plane details.
We also band-pass filtered the resulting maps to produce spatial plots of the
quadrupole, octopole and hexadacapole. 
We found that zeroing out the dirtiest parts of the Galactic plane 
had a negligible effect on both the power spectrum and on
the spatial structure of these lowest multipoles. Specifically, 
we could zero out all but the three cleanest regions \fig{masksFig}
(everything with $T>1$mK) without 
the quadrupole or octopole changing substantially. 
The spatial morphology of the quadrupole, octopole and hexadecapole 
for the all-sky analysis agrees well between our map and the ILC map, again 
showing insensitivity to galaxy modeling details 
(in particular, the ILC map is likely to have
less contamination in the Galactic plane due to more subdivisions there).

Although more detailed foreground modeling would be needed to rigorously quantify the
foreground contribution to low multipoles, let us, encouraged by the above-mentioned tests, 
tentatively assume that this contribution
is unimportant and perform an all-sky analysis of our cleaned map.
The resulting power spectrum is shown in \fig{quadrupoleFig}, which is simply a blow-up of
the leftmost part of the previous figure, and the corresponding angular correlation function
is shown in \fig{corrfuncFig}. 

Table 1 summarizes the quadrupole and octopole results.
We see that although the quadrupole is still low, it
is not quite as low as that from the cut-sky WMAP team analysis of \cite{hinshawpower03}.
Moreover, our map has a quadrupole virtually identical to the WMAP team ILC map despite
the differences in foreground modeling, further
supporting our hypothesis that the quadrupole is not strongly affected by foregrounds.
The second column in Table 1 shows the probability of the quadrupole in our Hubble volume
being as low as observed if the best-fit WMAP team model from 
\cite{Spergel03,verde03} is correct. This is computed for the all-sky case where 
$\delta T^2_2$ has a $\chi^2$-distribution with 5 degrees of freedom, so the 
probability tabulated is simply $1 - \gamma[5/2,(5/2) T^2_2/855.6\uK^2]/\Gamma(5/2)$, 
where $\gamma$ and $\Gamma$ are the incomplete and complete Gamma functions, respectively\footnote{
The actual probability is slightly larger for the cut-sky case where there
are fewer effective degrees of freedom, although \cite{Spergel03} show with Monte Carlo simulations
that there it is still a disturbingly small for the measured WMAP quadrupole --- indeed, they
obtain a consistency probability of $0.7\%$ using Markov chains.}.
We see that the statistical significance of the low quadrupole problem drops substantially
with our all-sky analysis, below the 95\% significance level, 
requiring merely a one-in-twenty fluke. 

To understand why the all-sky quadrupole is larger than the cut-sky one measured by the
WMAP team, we plot the lowest three multipoles
of our cleaned map in \fig{multipolesFig}, all on the same temperature scale.
Several features are noteworthy.
First of all, the quadrupole is low with an interesting alignment. 
A generic quadrupole has three orthogonal pairs of extrema
(two maxima, two minima and two saddle points). We see that 
the actual CMB quadrupole has its strongest pair of lobes, apparently coincidentally, 
fall near the Galactic plane.
Applying a Galaxy cut therefore removes a substantial fraction of the quadrupole power.
The saddle point is seen to be close to zero.
In other words, there is a preferred axis in space along which the observed quadrupole has almost no power.

Second, the observed quadrupole is the sum of the cosmic quadrupole and the dynamic 
quadrupole due to our motion relative to the CMB rest frame - see \cite{KamionKnox02} for a detailed
discussion. Since this motion is accurately known from the CMB dipole measurement 
\cite{Smoot92,BennettWMAP}, the dynamic quadrupole can and should be subtracted 
when studying the cosmic contribution. \Fig{multipolesFig} shows 
the dynamic quadrupole approximated by 
$(v/c)^2 (\cos\theta^2 - 1/3)$, where $v\approx 369 km/s$ towards
$(l,b)\approx (264,48)$ is the velocity of the
Solar System relative to the CMB \cite{BennettWMAP} and 
$\theta$ is the angle relative to this velocity vector.
This is a small correction with peak-to-peak amplitude 
$(v/c)^2\sim 4\uK$, and Table 1 shows that it reduces the cosmic quadrupole slightly.
This approximation is crude since the dynamic quadrupole in fact depends on frequency 
\cite{KamionKnox02}, and therefore may have been either over- or under-estimated in our foreground cleaned map
- we have shown it here merely to illustrate its spatial orientation and give an order-of-magnitude
estimate of its importance.

Third, although the overall octopole power is large, not suppressed like the quadrupole, 
it too displays the unusual property of a preferred axis along which power is suppressed.
Moreover, this axis is seen to be approximately aligned with that for the quadrupole.
The reason that our measured octopole in \fig{quadrupoleFig} is larger than that
reported by the WMAP team is therefore, once again, that much of the power falls 
within the Galaxy cut.
In contrast, the hexadecapole is seen to exhibit the more generic behavior we
expect of an isotropic random field, with no obvious preferred axis.

How significant is this quadrupole-octopole alignment?
As a simple definition of preferred axis for an arbitrary sky map, \cite{smalluniverse}
computes
the unit vector $\rh$ around which the angular momentum dispersion
$\expec{(\rh\cdot\L)^2}=\sum_m m^2 |a_{\l m}(\rh)|^2$
is maximized.
Here $a_{\l m}(\rh)$ denotes the spherical harmonic 
coefficients of the map in a rotated coordinate system with its 
$z$-axis in the the $\rh$-direction. 
The preferred axes $\rh_2$ and $\rh_3$ for 
the quadrupole and octopole, respectively, are
\beqa{axisEq}
	\rh_2&=& (-0.1145, -0.5265, 0.8424), \nonumber\\
	\rh_3&=& (-0.2578, -0.4207, 0.8698), 
\eeqa
\ie, both roughly in the direction of $(l,b)\sim (-110^\circ,60^\circ)$ in Virgo.
The angle between these two axes is merely $10.3^{\circ}$, and
the probability that a random axis falls inside a circle 
of radius $10.3^{\circ}$ around the quadrupole axis is simply the area 
of this circle over the area of the half-sphere, about $1/62$.
In other words, if the CMB is an isotropic Gaussian random field, 
then a chance alignment this good requires about a 1-in-62 fluke.
This issue is discussed in greater detail in \cite{smalluniverse}.
 
What does this all mean? Although we have presented these low multipole results 
merely in an exploratory spirit, and more thorough modeling of the foreground 
contribution to $\l=2$ and $\l=3$ is certainly warranted, 
it is difficult not to be intrigued by the similarities of \fig{quadrupoleFig}
with what is expected in some non-standard models, for instance ones involving 
a flat ``small Universe'' with a compact topology as in \cite{Starobinsky,Stevens,bb1,bb2,Levin01,Rocha}
and one of the three dimensions being relatively small (of order the Horizon size or smaller).
This could have the effect of suppressing the large-scale power in this particular
spatial direction in the same sense as is seen in \fig{quadrupoleFig}.
As so often in science when measurements are improved,
WMAP has answered old questions and raised new ones.

\bigskip

We thank the the WMAP team for producing such a superb
data set and for promptly making it public via
the Legacy Archive for Microwave Background Data 
Analysis (LAMBDA)\footnote{The WMAP 
data are available from
\protect\url{http://lambda.gsfc.nasa.gov}.
}.
Support for LAMBDA is provided by the NASA Office of Space Science.
We thank Krzysztof G\'orski and collaborators for creating the 
HEALPix package \cite{healpix1,healpix2},
which we used both for harmonic transforms and map plotting.
Thanks to Ed Bertschinger, Gary Hinshaw, Jim Peebles, Ned Wright and Matias Zaldarriaga for helpful comments.
This work was supported by 
NSF grants AST-0071213, AST-0134999, \& AST-0205981, and by
NASA grants NAG5-10763 \& NAG5-11099.
MT was supported by a David and Lucile Packard Foundation fellowship
and a Cottrell Scholarship from Research Corporation.



\end{document}